# ESA-ESO Working Groups
## The Herschel-ALMA Synergies

Chair: T. L. Wilson (ESO)
Co-chair: D. Elbaz (CEA)

# Report by the ESA-ESO Working Group on The Herschel-ALMA Synergies

## Introduction & Background

Following an agreement to cooperate on science planning issues, the executives of the European Southern Observatory (ESO) and the European Space Agency (ESA) Science Programme and representatives of their science advisory structures have met to share information and to identify potential synergies within their future projects. The agreement arose from their joint founding membership of EIROforum (http://www.eiroforum.org) and a recognition that, as pan-European organisations, they served essentially the same scientific community.

At a meeting at ESO in Garching during September 2003, it was agreed to establish a number of working groups that would be given the task of exploring these synergies in important areas of mutual interest and to make recommendations to both organisations. The chair and co-chair of each group were to be chosen by the executives but thereafter, the groups would be free to select their membership and to act independently of the sponsoring organisations.

## Terms of Reference and Composition

The goals set for the working group were to provide:

- A survey of the scientific areas covered by the Herschel and ALMA missions. This survey will comprise:
  a. a review of the methods used in those areas;
  b. a survey of the use of ALMA and Herschel in those areas and how these missions compete with or complement each other;
  c. for each area, a summary of the potential targets, accuracy and sensitivity limits, and scientific capabilities and limitations.

- An examination of the role of ESO and ESA in running these facilities that will:
  a. identify areas in which current and planned ESA and ESO facilities will contribute to science areas covered by Herschel and ALMA;
  b. analyse the expected scientific returns and risks of each;
  c. identify areas of potential scientific overlap, and thus assess the extent to which the facilities complement or compete;
  d. identify open areas which merit attention by one or both organisations (for example, follow-up observations by ESO to maximise the return from Herschel);
  e. conclude on the scientific case for coordination of ALMA and Herschel activities.

Edited by T. L. Wilson (ESO) and D. Elbaz (CEA)

The contributors to the scientific content of this report are: P. Andreani (Trieste), D. Bockelee-Morvan (Paris), J. Cernicharo (Madrid), P. Cox (Grenoble), C. De Breuck (ESO), E. van Dishoeck (Leiden), D. Elbaz (CEA, Saclay), M. Gerin (Paris), R. Laing (ESO), E. Lellouch (CEA, Saclay), G. L. Pilbratt (ESA), P. Schilke (Bonn), T. L. Wilson and M. Zwaan (ESO).

R. Fosbury and W. Freudling (ST-ECF) provided coordination, A. Rhodes and J. Walsh proof-read the final version of this document. Technical production was done by R. Y. Shida (ESA/Hubble), M. Kornmesser (ESA/Hubble) and L. L. Christensen (ESA/Hubble).

W. Fusshoeller (MPIfR Bonn), E. Janssen (ESO), J. Vernet (ESO) and W. Freudling provided figures.

# Contents

# 1 Executive Summary

The Herschel Satellite and the Atacama Large Millimeter Array (ALMA) are two very large sub-mm and far infrared (FIR) astronomy projects that are expected to come into operation in this decade. This report contains descriptions of these instruments, emphasising the overlaps in wavelength range and additional complementarities.

A short rationale for studying sub-mm and far infrared astronomy is given. Following this, brief presentations of Herschel and ALMA are presented, with references to more detailed documents and use cases. Emphasis is placed on the synergies between these facilities, and the challenges of comparing data produced using both. Specific examples of projects are given for a number of areas of astronomical research where these facilities will lead to dramatic improvements.

This report is addressed to an audience of non-specialist astronomers who may be interested in extending their areas of research by making use of Herschel and ALMA instruments. ALMA has a small instantaneous field of view, but allows high angular resolution images of selected sources. The Herschel satellite has two multi-beam bolometer systems, PACS and SPIRE. These have larger fields of view than ALMA, but with lower angular resolutions. Thus the SPIRE and PACS cameras provide the opportunity to cover large areas of the sky rather quickly. Measurements with ALMA would be follow-ups, while Herschel SPIRE and PACS can provide finding lists for ALMA, or for shorter wavelength measurements of source emission to give complete Spectral Energy Distributions (SEDs). Since Herschel will be above the atmosphere, measurements can be made in wavelength regions where astronomical signals cannot reach the Earth's surface. This is especially the case for wavelengths shorter than 300 µm. Herschel HIFI is a heterodyne instrument, so is especially well suited to high resolution spectroscopy of molecules such as water vapour. The number of sources that can be measured with HIFI will be more limited than PACS and SPIRE since HIFI is a single pixel instrument. On the Earth's surface, most of the water vapour lines are blocked by the atmosphere. The few that do reach the surface are nearly all strong masers. One exception seems to be the water vapour line at 183 GHz. Six antennas of ALMA will be equipped with receivers to image the $3_{13}$-$2_{20}$ transition of water vapour at 183 GHz and the same transition from the oxygen-18 isotope of water vapour at 203 GHz. Such high angular resolution images will complement the Herschel HIFI data.

A particularly important condition for combining ALMA and Herschel data involves a common source sample and consistent calibration. This calibration programme will require a fairly extensive set of Herschel and ALMA measurements, in addition to accurate models of the calibration sources. These sources will have to be more compact planets in the outer part of the Solar System or asteroids. For comparisons with ALMA, PACS and SPIRE calibrations will be more complex than HIFI calibrations.

Operationally, Herschel should take the lead in initiating projects. For extragalactic sources, PACS and SPIRE could survey large regions, providing finding surveys for ALMA. The ALMA follow-ups could be redshift determinations, an extension of the SED's to longer wavelengths, and high resolution imaging. For an efficient synergy, a significant amount of Herschel time should be devoted to legacy projects early in the life of the satellite. For galactic sources, sensitivity is less critical. There could be very large-scale surveys or deep targeted surveys of selected objects. Before full ALMA operation, ESO should consider carrying out large, very deep surveys with ground-based sub-mm single dishes equipped with large bolometer cameras.

For ESA, the most pressing need is for a coordination of large surveys planned in the Herschel guaranteed time. This includes both large scale and targeted surveys. ESA must have a clear policy of data rights. Ideally ESA should provide access to the Herschel data during the satellite's lifetime. This includes data files, calibration information and a pipeline for data reduction.

For both organisations, the most important task is to coordinate the large surveys that are planned in the Herschel guaranteed time by means of dedicated conferences or workshops.

# 2     Introduction

The Herschel satellite and the Atacama Large Millimeter Array (ALMA) are two large astronomical instruments designed to explore the "cool Universe", that is, to study cool gas and dust. Such cool material is associated with objects in formation or dust enshrouded sources. These include the earliest evolutionary stages of galaxies, stars and planets; these are deeply hidden within dust clouds where optical extinction can be extremely large. At far infrared and sub-mm wavelengths, the extinction is not only much smaller, but we can also directly measure the physical phenomena associated with the formation process itself.

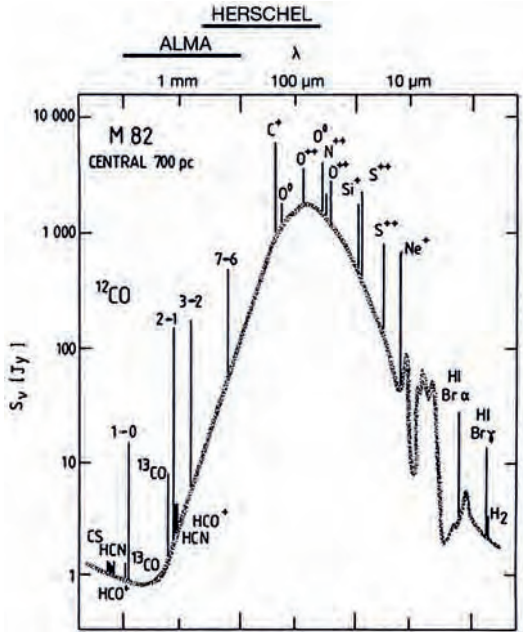

**Figure 1:** The Spectral Energy Distribution (SED) of a nearby starburst galaxy, M82 (from Genzel, 1991). Above the SED, we show the total wavelength and frequency ranges covered by the Herschel and ALMA instruments (Band 3 to 10, see Fig. 2).

Herschel will cover the wavelength ranges from 60 to 625 μm where the broadband dust emission peaks, allowing one to measure total bolometric luminosities and to characterise dust temperatures. ALMA will ultimately cover the range from 320 μm (~950 GHz) to 1 cm (~30 GHz) and will nearly always measure radiation from the optically thin Rayleigh-Jeans part of the spectrum and so can provide measurements of dust masses. Using assumed dust-to-gas ratios, it is possible to use dust masses to determine the total mass of the interstellar medium. With ALMA one can measure transitions of molecules with permanent dipole moments. In the sub-mm wavelength range, emission lines arise from gas with densities higher than those emitted at longer wavelengths. These results can be used to estimate molecular abundances and $H_2$ densities. A typical Spectral Energy Distribution (SED) is shown in Fig. 1. We also show the operational wavelength ranges of the two instruments.

Both Herschel and ALMA will come into operation at similar times. ALMA should be completed in 2012, but "early science" operation will begin well before this. The official launch date of the Herschel satellite is early 2008 and it has an expected lifetime of more than 3 years. Thus there should be a period during which both are in operation together. This will allow for the simultaneous measurement of astronomical sources with rapid time variations. However, even if the measurements do not overlap, one can combine results from these facilities to improve the analysis of sources with slow or no time variations.

Hence it is important to establish regions of common ground and to develop strategies that will optimise the opportunities to exploit the distinctive features of each instrument and define combined programmes aimed at specific science goals. This report reviews the capabilities of each instrument and compares their performance and approach to common observational techniques in Chapters 3 and 4. If the observational areas of the two instruments are to overlap, some form of common calibration is desirable and methods for this are discussed in Chapter 5, before specific examples of potential common science programmes are given in Chapter 6. Recommendations for the future, addressed separately and jointly to ESO and ESA, are presented in Chapter 8.

# 3 Descriptions of Herschel and ALMA in the Literature

A description of the bilateral (North America-Europe) ALMA project is at http://www.alma.nrao.edu/projectbk/construction/. Accounts of ALMA science are in Wootten (2001) and Shaver (1995). The website for the Herschel project, including all instruments, is http://www.rssd.esa.int/Herschel/. In particular, descriptions of the science that will be carried out by Herschel/SPIRE (Spectral & Photometric Imaging Receiver) are found in the SPIRE web page: http://www.ssd.rl.ac.uk/SPIRE/Science.htm, with PACS (Photodetector Array Camera & Spectrometer) in the PACS web page: http://pacs.ster.kuleuven.ac.be/
and for the Herschel HIFI in the web page: http://www.sron.nl/divisions/lea/hifi

Accounts of Herschel and ALMA, some plans for Herschel science, ALMA science and their synergies are to be found in the *Proceedings of The Dusty and Molecular Universe* (ed. A. Wilson, 2005). The *Molecular Universe Research Training Network* is involved directly in Herschel projects, as well as laboratory measurements and analysis tools to support Herschel observing projects. A description of the network, including links to Herschel activities is at http://molecular-universe.obspm.fr/.

## 3.1 The ALMA Project

The characteristics of ALMA are determined by three primary science goals. The first two are specific science cases. The first is detecting a Milky Way type galaxy at redshift z = 3, and the second is imaging planet-forming regions in a protostellar disk at 140 pc, the distance to the nearest star-forming regions. These determine the sensitivity and highest angular resolution of ALMA. The third requirement demands that ALMA produce high quality images at high spatial resolutions.

In the millimetre and sub-mm wavelength range, the Earth's atmosphere has a significant influence on the measurements of astronomical sources. To minimise this influence, ALMA is located on a dry site at an elevation of 5 km. A sketch of the frequency coverage of the planned set of receivers is shown in Fig. 2. Parameters of ALMA as known at the end of 2005 are given in Tables 1 and 2. In the bilateral ALMA baseline plan, receiver bands 3, 6, 7 and 9 will be provided.

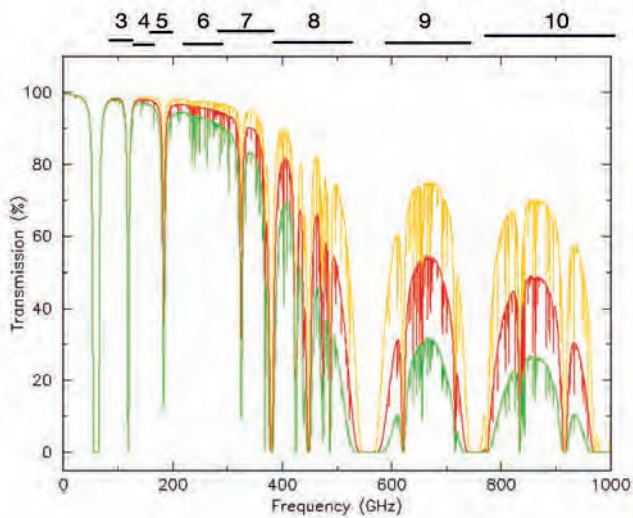

**Figure 2:** The numbered horizontal lines at the top show the coverage of ALMA receiver bands 3 to 10. Receiver bands 3, 6, 7 and 9 are in the North American-European bilateral plan. Receiver bands 4, 8 are to be contributed by Japan. There are plans for Japan to provide Band 10, dependent on R & D work. Bands 1 and 2 are not yet planned. Six receivers for Band 5 will be provided by EC funding in FP6. At any time, observations can be carried out with only one receiver band. The coloured curves shows three atmospheric transmissions for total perceptible water vapour (pwv) amounts of 0.2, 0.5 and 1 mm for Mauna Kea (from http://www.submm.caltech.edu/cso/weather/atplot.shtml).

It is now certain that Japan will enter the ALMA project. Receiver Bands 4, 8 will also be provided by Japan. There are plans for Japan to provide Band 10 eventually, after further research and development. Japan will also provide the "ALMA Compact Array" (ACA) that consists of twelve 7 m antennas to provide measurements of more extended structure and four 12 m antennas, to be used in the single dish mode to measure the total flux density of a source. The parameters of the ALMA project given in Table 1 reflect the result of a rebaseline process for the bilateral array. At present 50 antennas are planned, with an increase to 64 antennas if funding allows. An interactive sensitivity calculator for all ALMA Bands except 1, 2 and 10 is at: http://www.eso.org/projects/alma/science/bin/sensitivity.html. This calculator takes into account the effect of the Earth's atmosphere.

**Table 1:** ALMA Antenna Arrays and Configurations

| Array | Bilateral | Compact (ACA) |
|---|---|---|
| Number of Antennas | 50, up to 64 | 16 |
| Total Collecting Area | 5654-7237 m² | 915 m² |
| Array Configurations (dimension of region filled) Compact filled Largest extent | 150 m 18.5 km | 35 m |
| Total Number of Antenna stations | 175 | 22 |
| Antenna Diameter | 12 m[a] | 4 x 12 m[b]; 12 x 7 m[a] |
| Surface accuracy | 25 μm | 12 m: 25 μm; 7 m: 20 μm |
| Field of View | | |
| 3mm | 50" | 50"; 85" |
| 0.5 mm | 8.3" | 8.3"; 14" |

[a] Transportable by especially constructed vehicles
[b] Fixed in position

**Table 2:** ALMA Front Ends [a]

| Band 1 | (1cm) | 31.3-45 GHz | HEMT[b] |
|---|---|---|---|
| Band 2 | (4mm) | 67-90 GHz | HEMT |
| Band 3 | (3mm) | 84-116 GHz | SIS [c,d] |
| Band 4 | (2mm) | 125-163 GHz | SIS [e] |
| Band 5 | (1.8mm) | 163-211 GHz | SIS [f] |
| Band 6 | (1.3mm) | 211-275 GHz | SIS [d] |
| Band 7 | (0.9mm) | 275-373 GHz | SIS [d] |
| Band 8 | (0.6mm) | 385-500 GHz | SIS [e] |
| Band 9 | (0.5mm) | 602-720 GHz | SIS [d] |
| Band 10 | (0.3mm) | 787-950 GHz | SIS [g] |

[a] All with dual polarisation; each polarisation feeds four IF sections, each with a bandwidth of 2 GHz, giving a total bandwidth of 8 GHz in each polarisation. Also uncooled 183 GHz water vapour monitors provide data for phase corrections
[b] High Electron Mobility Transistor receiver
[c] Superconductor-Insulator-Superconductor mixer receivers.
[d] In the bilateral baseline plan
[e] Contribution from Japan
[f] Funded by EC for 6 antennas
[g] Contribution from Japan; depends on R&D programme

The angular resolution is generally given by the expression

$$\theta = 0.2 \lambda / B$$

Here $\theta$ is in arc seconds, $\lambda$ is the observing wavelength in millimetres and B is the largest separation of the antennas in km. For $\lambda = 3.5$ mm and B = 0.15 km, $\theta = 4"$ while for $\lambda = 0.5$ mm, and B = 14.5 km, $\theta = 0.005"$. This relation also allows a calculation of the field of view (FOV) of ALMA, if one uses the diameter of the antenna for the term B. For a 12 m antenna at 3.5 mm, the FOV will be 58" while for the ACA 7 metre antennas the FOV is 100".

For each receiver band, there are four intermediate frequency (IF) sub-bands, each with a total bandwidth of 2 GHz. In each, the ALMA spectrometer provides a range of velocity resolutions. The lowest resolution is 128 channels of 18.9 MHz, covering 2 GHz for each of the 2 polarisations. The spectrometer has the property that the product of total bandwidth (in GHz) with the number of channels is 256. These spectrometer windows can be placed nearly anywhere in each 2 GHz wide IF sub-band. For spectroscopic surveys one can measure 4 sub-bands of a given velocity resolution simultaneously within an 8 GHz wide region. The finest resolution will be 3.1 kHz; this corresponds to 0.01 kms$^{-1}$ at 100 GHz, the middle of receiver Band 3. This velocity resolution corresponds to 10% of the expected collapse speed of a molecular cloud that will form a low mass star.

Since ALMA is made up of individual antennas, it is possible, indeed desirable, to begin "early science" measurements before full completion in 2012. In the early science mode, at least 6 antennas will be available. Thus the ALMA images will be less detailed compared to those possible with the full ALMA and the sensitivity will be about a factor of 8 lower.

Examples of ALMA scientific projects with a 64 antenna bilateral array can be found in the *Design Reference Science Plan* (DRSP) at http://www.strw.leidenuniv.nl/~alma/drsp.shtml. The DRSP contains examples of prototype projects to demonstrate the capabilities of ALMA. The DRSP consists of a mix of small and large programmes, written by researchers currently active in the field. Some of the larger programmes could evolve into "legacy" or "key programmes", but could also be carried out as a number of smaller programmes by different groups with essentially the same science goals but different sources. The DRSP contributions do not reserve research areas for the authors, but are meant to be examples to provide estimates of integration times, calibration accuracies and noise limits.

At longer wavelengths where the FOV is fairly large, ALMA can be used to carry out blind surveys, but since the field of view of the antennas is small, a more efficient approach would be to use ALMA for imaging selected sources. There are many methods to select such sources. One is to use the Atacama Pathfinder Experiment (APEX) 12 m sub-mm antenna to survey selected regions. APEX could be used to find star-forming regions in the galactic plane or survey high galactic latitude regions. Other instruments that could be used to provide finding lists are the James Clerk Maxwell (JCMT) 15 m sub-mm antenna on Mauna Kea or the 50 m Large Millimeter Telescope (LMT) in Mexico. The sources found would be measured in follow-up surveys with ALMA.

## 3.2 Herschel

Herschel is a cornerstone mission of the European Space Agency. A recent report on the Herschel mission has been given by Pilbratt (2005). The Herschel satellite will orbit at the Lagrange point L2, and will have three instruments mounted on a single 3.5 metre passively cooled antenna. The broadband instrument systems consist of the Photodetector Array Camera & Spectrometer, PACS (Poglitsch et al., 2005) and Bolometer cameras, the Spectral & Photometric Imaging Receiver, SPIRE (Griffin et al., 2005). The PACS instrument consists of photometers that can perform measurements at 75, 110 or 170 μm. We list some properties of the PACS broadband instrument in Table 3. The SPIRE instrument consists of cameras that can perform measurements at 250, 360 or 520 μm. We list some properties of the SPIRE instrument in Table 4. The PACS and SPIRE beamsizes are compared with an image of the Hubble Deep Field image in Fig. 3. This gives a qualitative estimate of source confusion due to the angular resolution of PACS and SPIRE. For very deep integrations at shorter wavelengths there should be little source confusion, but confusion will be a consideration at longer wavelengths (see section 6.1).

Both SPIRE and PACS can be used as low resolution spectrometers. The PACS spectrometer can be used for simultaneous measurements at 57-105 μm and 105-210 μm. See Table 5 for a list of some of the PACS spectrometer properties. The SPIRE spectrometer can perform measurements at 200-300 or 300-670 μm. We list a summary of the low resolution Herschel spectrometers in Table 6.

There is a single pixel heterodyne system, Heterodyne Instrument for the Far Infrared, Herschel HIFI (Graauw & Helmich, 2001). Herschel HIFI is a single pixel high resolution spectrometer.

**Table 3:** Some Characteristics of the PACS Photometer

| Central Wavelength (μm) | Angular Resolution (") | Number of pixels | Field of View (') |
|---|---|---|---|
| 75 | 5.4 | 64 by 32 | 1.75 by 3.5 |
| 110 | 8.0 | 64 by 32 | 1.75 by 3.5 |
| 170 | 12.2 | 32 by 16 | 1.75 by 3.5 |

**Table 4:** Some Characteristics of SPIRE

| Central Wavelength (μm) | Pixel Size (") | Angular Resolution (") | Number of pixels | Field of View (') |
|---|---|---|---|---|
| 250 | 12 | 17 | 139 | 4 by 8 |
| 360 | 18 | 24 | 88 | 4 by 8 |
| 520 | 24 | 35 | 43 | 4 by 8 |

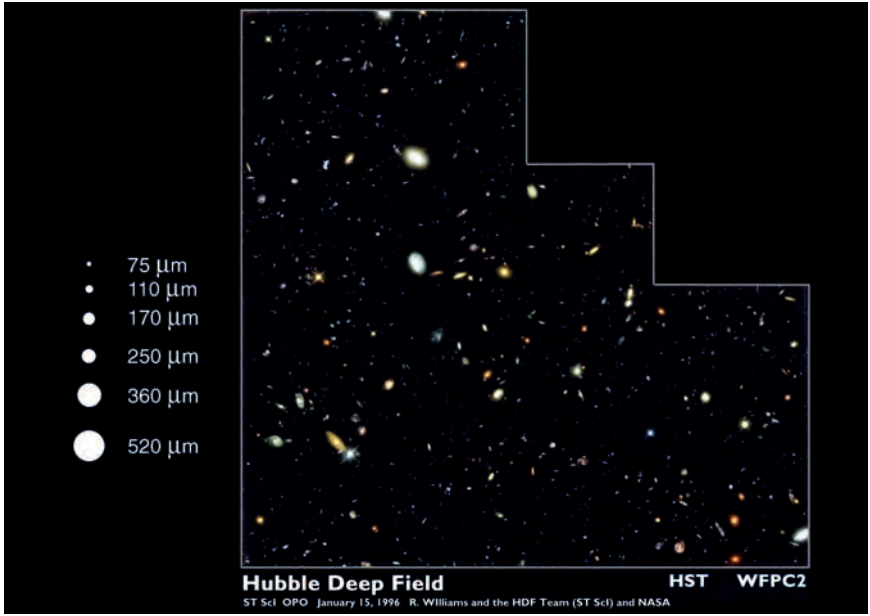

**Figure 3:** The Hubble Deep Field (angular resolution 0.05") is shown on the right. On the left are the Full Width to Half Power beamsizes of the PACS (three highest) and SPIRE instruments (three lowest).

**Table 5:** Some Characteristics of the PACS Spectrometer

| Central Wavelength (μm) | Angular Resolution (") | Number of pixels | Field of View (") | Spectral Resolution (approximate) |
|---|---|---|---|---|
| 72 | 9.4 | 16 x 25 | 47 by 47 | 1800 |
| 105 | 9.4 | 16 x 25 | 47 by 47 | 1000 |
| 210 | 9.4 | 16 x 25 | 47 by 47 | 2000 |

Since Herschel is above the atmosphere, signals are not blocked by terrestrial spectral features. This location is ideal for the measurement of water vapour lines that are usually absorbed by the atmosphere even at the ALMA site. Since the Herschel HIFI receivers are double sideband mixers, the receiver noise for spectral line measurements is twice the double sideband (DSB) system noise. To relate the measured temperatures to flux densities, there is a correction for antenna efficiency but no correction for the Earth's atmosphere (see caption of Fig. 4). Thus, the Herschel HIFI receiver noise temperatures, shown in Fig. 4, are all important for sensitivity. As with ALMA, the velocity resolution of Herschel HIFI can be extremely high. The finest spectrometer resolution is 140 kHz, corresponding to 0.1 kms$^{-1}$ at 480 GHz with finer values at higher frequencies.

**Table 6:** Some Characteristics of the Herschel Spectrometers

| System | Wavelength Limits (μm) | Resolution (kms$^{-1}$) | Spectral Coverage (kms$^{-1}$) | Field of View (') |
|---|---|---|---|---|
| PACS | 57-210 | 80-300 | 700-3000 | 0.8 |
| SPIRE | 200-300 | 300-15000 | 2500-125000 | 2.6 |
| SPIRE | 300-670 | 480-24000 | 4000-200000 | 2.6 |
| HIFI | 157-212 | 0.02-0.2 | 960-2500 | 0.2 |
| HIFI | 240-625 | 0.1-0.6 | 850-625 | 0.8 |

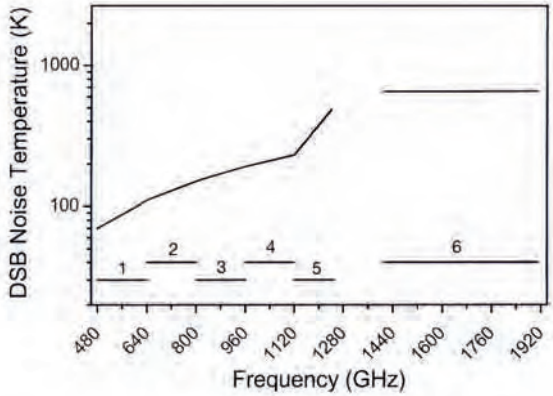

**Figure 4:** The Double Sideband (DSB) noise temperatures in Kelvin for Herschel HIFI receivers plotted versus sky frequency. The receiver bands are numbered above the horizontal lines at bottom of plot. For a 3.5 m antenna, the relation between flux density $S_\nu$ in Jy and antenna temperature, $T_A$, in K is $S_\nu = 287.3 T_A/\eta_A$, where the Herschel antenna efficiency is $\eta_A$. The finest frequency resolution of the Herschel HIFI spectrometer is 140 kHz, or 0.042 kms$^{-1}$ at 1 THz. The 1-sigma uncertainty for a 140 kHz spectrometer resolution is $\Delta T_{RMS} = 5.34 \times 10^{-3}\, T(DSB)\, Hz^{-1/2}$, Jackson (2005).

# 4   Comparison of the ALMA and Herschel

This section makes a brief comparison between the fundamental properties of ALMA and Herschel, demonstrating that they will have angular resolutions comparable with (ALMA) or approaching (Herschel) those of today's cutting edge telescopes, while operating at wavelengths not covered by current instruments. In addition, two principal observing strategies are presented.

## 4.1   Comparison of the Properties of ALMA and Herschel

In Fig. 5 we show the angular resolutions of the Very Large Array (VLA), ALMA, Herschel, the James Webb Space Telescope (JWST), the Very Large Telescope (VLT), the VLT Interferometer (VLTI) and the Hubble Space Telescope (HST). The range of angular resolutions for ALMA and the VLA correspond to the different configurations. This two-dimensional plot does not contain the frequency resolutions of the instruments. These are included in the projected three-dimensional illustration shown in Fig. 6.

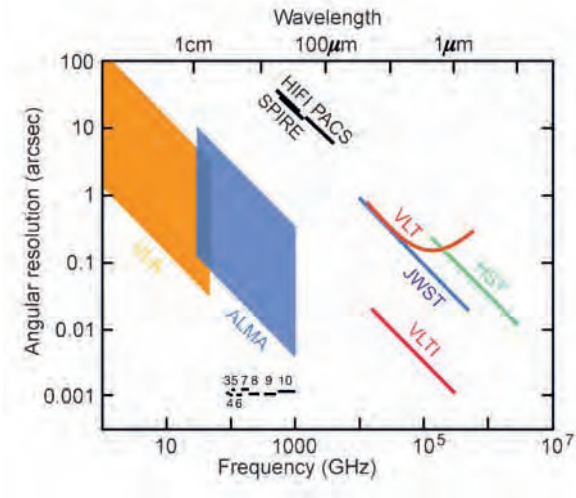

**Figure 5:** A plot of the angular resolutions of the VLA, ALMA, Herschel instruments, VLT, the VLT Interferometer, VLTI, JWST and HST. Below the region marked "ALMA" are the ALMA receiver bands.

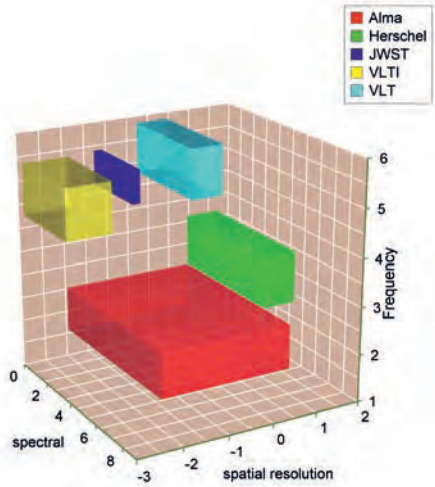

**Figure 6:** A three-dimensional plot of frequency, angular resolution and resolution for the instruments compared in Fig. 5.

From Fig. 6, we see that Herschel HIFI approaches the velocity resolution capable with ALMA, while for all other instruments, the velocity resolutions are much lower.

## 4.2 Comparisons of Observational Techniques

### 4.2.1 Broadband Measurements

ALMA will have a much larger collecting area than Herschel. However, at specific frequencies, usually near water vapour lines, the Earth's atmosphere attenuates signals, perhaps completely blocking these. For the SPIRE bolometer array, there is a significant overlap with ALMA receiver bands 8 to 10. The PACS bolometer array provides shorter wavelength data needed to determine SEDs. In Fig. 7 we show the coverage on a spectrum of the starburst galaxy M82 for different redshifts.

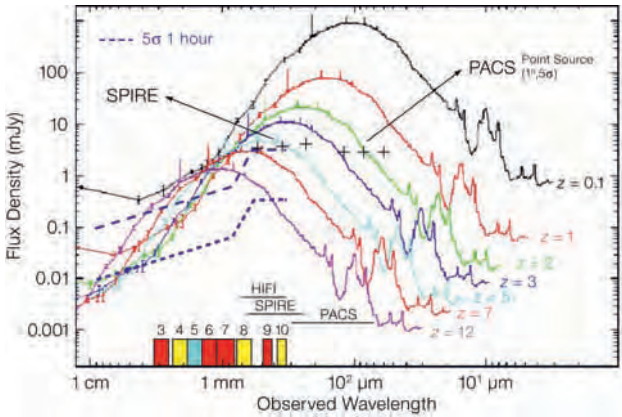

**Figure 7:** A plot of the emission from the starburst galaxy M82 for different redshifts, z. The horizontal axis is observed wavelength, the vertical axis is predicted flux density in mJy. The crosses show the sensitivity of the Herschel bolometers, taken from Griffin et al., 2005. The dashed lines in the left side of this diagram show the 5-sigma sensitivity of ALMA. The lower dashed curve is for the 64 antenna ALMA and the upper curve for a 6-antenna ALMA. The ALMA receiver bands are shown numbered above the horizontal axis. Cf Fig. 1 for line identifications. The coverages of the Herschel instruments are shown as horizontal lines (HIFI, SPIRES, PACS).

Even at redshifts of $3 < z < 5$, the broadband emission of sources such as M82 can be detected with Herschel and the early science ALMA.

To compare ALMA with Herschel data accurately, there are two important considerations. The first is the signal-to-noise ratio. From Fig. 7, we see that SPIRE and PACS will be able to obtain good measurements of M82-like sources for redshifts up to $z = 2$ in 1 hour. For higher redshifts, much longer integration times are needed. The second consideration is concerned with systematic effects. For example, the bandwidths of PACS and SPIRE detectors are much larger than the bandwidth of ALMA receivers. The most obvious requirement of a calibration plan must be to reduce or eliminate systematic effects such as the very different bandwidths and angular resolutions of these instruments (see below).

### 4.2.3 Spectral Line Measurements

The spatial distribution of a spectral line from a species is a result of excitation and chemistry. Given the distribution of a specific transition, one can model the distribution of other transitions given the temperature and density distributions. With Herschel, one can measure a large number of water vapour lines and with ALMA one can obtain high resolution images of many transitions simultaneously. To make convincing comparisons, it is desirable to measure the same line with both instruments.

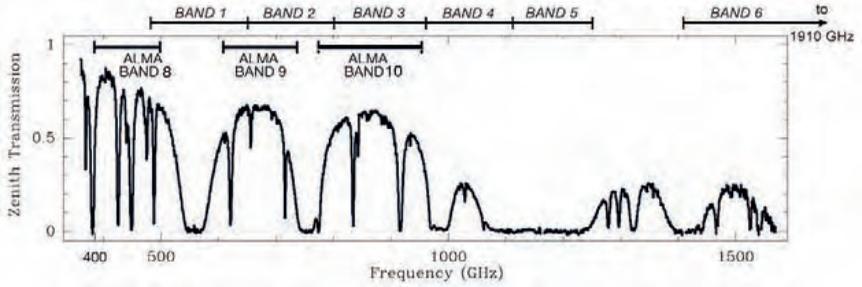

**Figure 8:** A plot of the Herschel HIFI receiver bands (above the plot) and ALMA receiver bands 8, 9 and 10. The irregular curve below the bands is the transmission at the zenith, as obtained from the data of Serabyn et al. (1998) for Mauna Kea.

With the inclusion of the contributions from Japan to ALMA, it will be possible to measure the same spectral lines in ALMA Bands 8, 9 and 10 and Herschel HIFI Bands 1 and 2 (see Fig. 8). In Table 7, we give a list of some of the spectral lines common to both instruments. In addition, there are thousands of spectral lines of methanol, $CH_3OH$, methyl cyanide, $CH_3CN$, sulphur monoxide, SO, and sulphur dioxide, $SO_2$ common to both instruments. For each spectral line, we give the quantum numbers of the transitions and line frequencies. We also give the atmospheric transmission for a 5 km high site with 0.5 mm of precipitable water (this is scaled to 5 km elevation from http://www.submm.caltech.edu/cso/weather/atplot.shtml).

**Table 7:** Frequencies, Assignments & Estimated Atmospheric Transmission where ALMA and Herschel HIFI Receiver Bands Overlap

| Molecule or Atom | Transition | Energy of Lower Level (K) | Frequency (GHz) | Transmission (percent) |
|---|---|---|---|---|
| ALMA Band 8: | | | | |
| CS | J = 10-9 | 106 | 489.8 | 37.1 |
| CI | $^3P_1$-$^3P_0$ | 0 | 492.2 | 51.8 |
| $NH_2D$ | $J_{KaKc} = 4_{13}$-$4_{14}$ | 152 | 494.4 | 54.3 |
| HDO | $J_{KaKc} = 1_{10}$-$1_{01}$ | 22 | 509.3 | 38.9 |
| ALMA Band 9: | | | | |
| HDO | $J_{KaKc} = 2_{11}$-$2_{02}$ | 66 | 599.9 | 21.7 |
| $D_2O$ | $J_{KaKc} = 3_{03}$-$1_{11}$ | 26 | 607.4 | 30.2 |
| $HCO^+$ | J = 7-6 | 90 | 624.4 | 21.3 |
| SiH | $J_F = (3/2)_1$-$(1/2)_1$ | 0 | 624.9,627.7 | 24.9,36.5 |
| $H^{37}Cl$ | J = 1-0 | 0 | 625.0,625.9 | 25.5,30.6 |
| DF | J = 1-0 | 0 | 651.1 | 51.1 |
| $^{13}CO$ | J = 6-5 | 79 | 661.0 | 53.9 |
| $D_2H^+$ | | | 690 | 49.9 |
| CO | J = 6-5 | 83 | 691.5 | 48.1 |
| ALMA Band 10: | | | | |
| CO | J = 7-6 | 116 | 806.6 | 39.2 |
| CI | $^3P_2$-$^3P_1$ | 24 | 809.3 | 40.6 |
| $H_2CO$ | $J_{KaKc} = 12_{0,11}$-$11_{0,11}$ | 228 | 855.2 | 48.9 |
| $^{13}CO$ | J = 8-7 | 148 | 898.0 | 40.6 |
| HDO | $J_{KaKc} = 1_{11}$-$0_{00}$ | 0 | 893.6 | 40.1 |
| LiH | J = 2-1 | 21 | 887.3 | 34.0 |
| $H_2CO$ | $J_{KaKc} = 12_{1,11}$-$11_{1,10}$ | 249 | 896.7 | 40.5 |
| $D_2O$ | $J_{KaKc} = 2_{12}$-$1_{01}$ | 17 | 897.9 | 40.6 |

# 5 Common Calibration of ALMA and Herschel data

By common calibration we mean measurements of the same sources using ALMA and Herschel. These sources must have known continuum spectral indices since the bandwidths of PACS and SPIRE are much larger than those of ALMA receivers. In many cases one must rely on accurate models of sources. In addition, one may have to account for luminosity variations and the presence of narrowband features in the spectra of calibration sources. The Herschel spacecraft will not be permitted to point close to the Sun. The PACS and SPIRE detectors will saturate when measuring intense sources and these will be confusion limited when measuring faint sources. Thus, for the faint flux density ranges, calibration sources might have to be asteroids (e.g., Ceres, Pallas, Vesta, Hygiea), outer planets (Pluto/Charon system) and/or moons of the giant planets. If higher flux densities are needed, Mars and the giant planets must be used in both lines and continuum. The brightness temperature in all the observed bands must be established from model fitting and observations of frequency ranges free from atmospheric lines (water and CO for Mars) and broad spectral features (for giant planets). The surface emissivity and expected solid-state features must be accurately known.

For Mars, in addition to the rocky surface, one must model the influence of dust storms, the ice caps and the seasonal variation of the brightness temperature. For giant planets, brightness, temperature and density of the atmosphere, and the abundance profiles and opacities in lines and continuum must be carefully analysed. The calibrations must be based on a comprehensive comparison between models and observations, clearly defining the model assumptions. This will require careful planning of observations, both from ground-based and from space facilities (eg, the ASTRO-F and Spitzer satellites) before Herschel and ALMA measurements can begin.

For spectral line calibrations, those transitions common to both instruments are listed in Table 7. The Herschel beamsize is larger, so for extended sources ALMA must be used to map the Herschel beam.

More about the proposed strategies for calibrating the Herschel instruments may be found in http://www.rssd.esa.int/Herschel/hcal_wkshop.shtml

## 5.1 Data Taking Procedures

These procedures are rather different for Herschel and ALMA. Herschel broadband data will be taken, employing chopping to balance the emission from the passively cooled telescope. In contrast, ALMA data taking involves the correlation of outputs of different antennas, supplemented by measurements of the source total flux densities with four 12 m single dishes. To take broadband data with these single dishes, ALMA must use wobbler switching to balance the emission from the Earth's atmosphere. Thus the responses of both instruments are biased against extended continuum emission. This may lead to systematic differences in Herschel and ALMA data caused by the different chopper throws. In addition, unlike Herschel, ALMA data must in any event be corrected for absorption by the Earth's atmosphere.

The PACS and SPIRE bolometer systems on Herschel are well suited to surveys of large regions of the sky. For a reasonable amount of time, ALMA can provide high sensitivity, high angular resolution images in spectral lines and continuum of regions up to a few arc minutes in size. Thus, ALMA is better suited to be a follow-up instrument for Herschel surveys.

Since the Herschel measurements are carried out from above the atmosphere, these measurements would be unaffected by the atmosphere. For this reason, the calibration should be much more stable.

## 5.2 Qualitative Summary of the Comparison

ALMA can provide higher angular resolution than Herschel HIFI, but the atmosphere blocks a significant part of the millimetre and sub-mm spectrum, including nearly all of the water vapour lines. For this reason, most water lines are not included in ALMA receiver bands. A significant exception is ALMA Band 5 which includes the $3_{13}$-$2_{20}$ transition of $H_2O$ at 183 GHz and $H_2^{18}O$ at 203 GHz. There will be 6 antennas equipped with Band 5 receivers. Measurements at the 183 GHz spectral line of water vapour are possible in the best 20% of the time on the ALMA site for galactic sources. For sources with significant redshifts, the Earth's atmosphere is less important, but sensitivity may play a role. Measurements of these spectral lines will supplement the lower angular resolution Herschel HIFI measurements of more than 30 lines of water vapour.

# 6 Specific Examples of Synergies Between ALMA and Herschel

In the following sections, we give eight examples of synergies. These are concerned with different categories of sources that are commonly measured.

## 6.1 Herschel and the origin of the cosmic infrared background (the issue of the confusion limit)

The cosmic infrared background (CIRB) is the product of the light radiated by dust over a Hubble time. Energetic considerations suggest that the CIRB is primarily powered by star formation with a contribution of less than 20% from accretion around a black hole (Fadda et al., 2002). The integrated cosmic infrared background (CIRB) is at least as large, to within the uncertainties, as the UV-visible-near infrared background (COB, cosmic optical background), hence it is a fossil memory of the past activity of distant galaxies. While mid infrared (MIR) observations with ISO (Elbaz & Cesarsky, 2003) and Spitzer (Dole et al., 2006) and sub-mm surveys with SCUBA (Barger et al., 1999) have revealed a population of galaxies that could produce the CIRB, its origin will only be fully confirmed after actually detecting those galaxies responsible for the 100-1000 μm radiation. Herschel offers a unique opportunity to address this issue, but is limited by spatial resolution. This will lead to source confusion that prevents one from resolving the CIRB into individual galaxies for wavelengths longward of 100 μm where the most distant galaxies are expected to emit. The Spitzer Space Telescope can operate at shorter wavelengths than Herschel. The Spitzer view of the MIR offers a complementary, but limited view of the distant dusty Universe, since the k-correction (Fig. 9) discriminates against detections of galaxies more distant than z~3. The 24 μm band of the MIPS photometer of Spitzer is shifted outside of the dust emission peak at larger redshifts. By imaging individual Herschel sources, ALMA will allow a separation of adjacent galaxies. Then it can be determined whether a Herschel source is the combination of several galaxies. Thus a combination of ALMA and MIR data will allow great progress in the study of the CIRB. While galaxies detected at 100-500 μm should be detectable with ALMA, ALMA by itself will not provide a robust determination of the emission at shorter wavelengths. ALMA's small field of view will prevent it from covering very large regions of the sky. However, even though ALMA measurements will be limited to relatively few sources initially, with time the ALMA sample will become larger and it will be possible to obtain high resolution images of thousands of Herschel sources. Ultimately, surveys with future infrared interferometers may be needed for a complete picture. However the combination of Herschel and ALMA will provide an excellent first cut at understanding the origin of the CIRB.

Table 8 shows the completeness of the six passbands of Herschel in regard to the CIRB. Column 1 contains the name of the instrument and wavelength, Column 2 gives the sensitivity limits determined by source confusion, and Column 3 contains the percentage number of sources that can be resolved at the confusion limit. The range of values represents the predictions for different models.

**Table 8**: Model Predictions for Herschel Surveys

| λ(μm)     | Depth (mJy) | % CIRB |
|-----------|-------------|--------|
| PACS-75   | 0.2 – 0.7   | 95-100 |
| PACS-110  | 1.5 – 5.0   | 86     |
| PACS-170  | 9.0 – 20.0  | 58     |
| SPIRE-250 | 22 - 42     | 22     |
| SPIRE-350 | 27 - 52     | 7      |
| SPIRE-500 | 25 - 39     | 2      |

In summary, for high redshift objects, Herschel will be confusion limited at the long wavelengths. It may be possible to lower the long wavelength confusion limit by deconvolving the Herschel beam with a point spread function that is well determined. At the shorter wavelengths, the confusion is less, but the k-correction is not so favourable, so the redshift limit, due to sensitivity, will be z = 2-5.

The study of the CIRB is a good example of the synergy between Herschel and ALMA. By combining the ALMA high resolution images with the shorter wavelength Herschel data, one can advance futher than with only one of these instruments.

## 6.2 Galaxy and large-scale structure formation (the issue of the size of extragalactic surveys)

Herschel is well suited for very large scale surveys, so one can study the effect of environment on the star formation rate and the infrared emission in specific galaxies. The most luminous infrared galaxies, powered either by star formation or an active galactic nucleus (AGN), are triggered by neighbouring galaxies. Moreover, the most massive galaxies, which are also expected to have formed through the strongest episodes of star formation and AGN activity, are located in the centre of galaxy clusters or, at least, in environments denser than those of less massive galaxies. By covering regions of the sky larger than the typical scale of large-scale structures, i.e. ~40-100 Mpc (in the co-moving frame) corresponding to ~40-100 arc-minutes at z~1 and 20-50 arc-minutes at z~3, Herschel will probe the epoch when the formation of large-scale structures influenced that of galaxies.

One expects that several square degrees must be surveyed around a specific source to check for such effects, and to potentially include a massive cluster of galaxies. Such surveys are outside of the scope of ALMA, but these surveys will benefit from the ALMA follow-ups for selected targets.

## 6.3 High Redshift Objects

During the past decade, our knowledge of high redshift (z > 1) galaxies and active galactic nuclei has substantially improved. One of the breakthroughs has been the study of their thermal dust and molecular line emission. The molecular lines, which are strongest at mm and sub-mm wavelengths, are excellent tracers of the reservoirs feeding such starburst objects. Millimetre and sub-mm observations are possible thanks to the development of sensitive bolometer arrays such as SCUBA on the JCMT (Holland et al., 1999), and MAMBO on the IRAM 30 m telescope (Kreysa et al., 1998). In addition, there are additional instruments, namely millimetre interferometers with large collecting area and more sensitive line receivers such as the IRAM Plateau de Bure facility. Observations of high redshift objects at millimetre wavelengths greatly benefit from the fact that the bulk of the luminosity is shifted into the millimetre and sub-mm range. This "negative k-correction" leads to an almost constant flux density from z = 1 to 10 (see Figure 9).

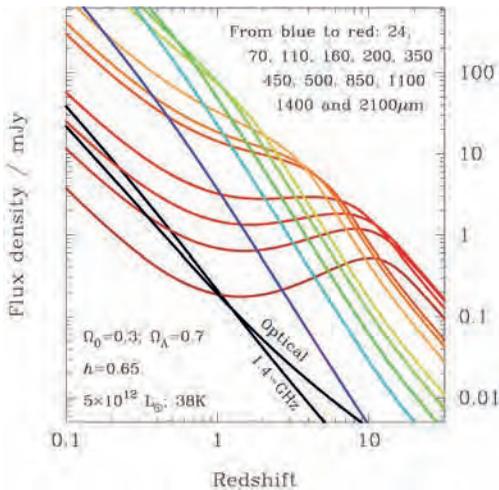

**Figure 9:** The predicted flux density of a dusty galaxy with a luminosity of 5 x $10^{12}$ solar luminosities at a temperature of 38 K is shown as a function of redshift for various wavelengths. The k-correction (the shift of luminosity with redshift) was calculated for the parameters of the Universe shown in the lower left of the figure. This k-correction favours detection of sources at wavelengths longer than about 250 μm, since the flux densities are almost independent of redshift. Reproduced from Blain et al. (2002).

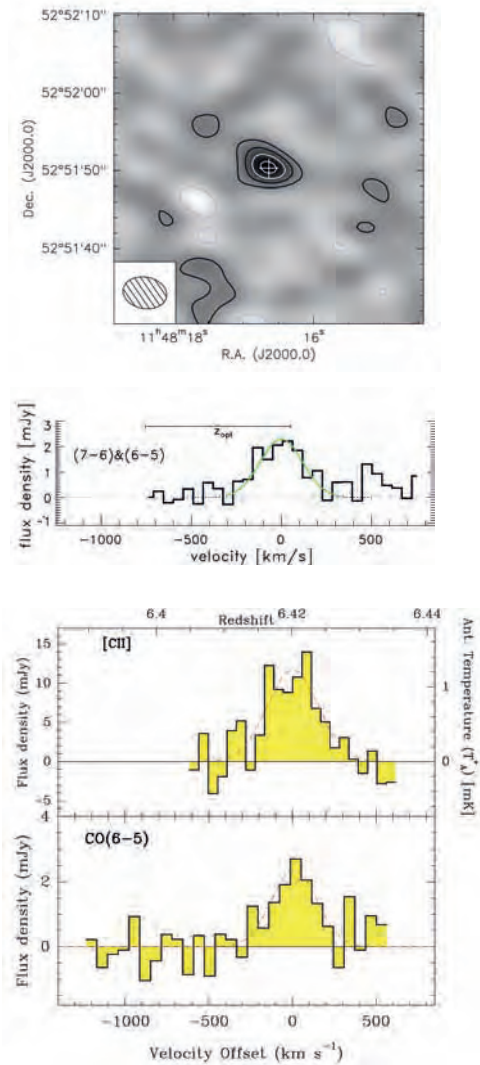

**Figure 10:** At the top, a velocity integrated image (v = -227 to 213 kms$^{-1}$) of the redshift z = 6.4 QSO (J114816.64+525150.3) obtained with the IRAM Plateau de Bure interferometer. In the middle the combined spectrum of the J = 6-5 & 7-6 lines of CO. At the bottom is the CII detection and the J = 6-5 line of CO (Maiolino et al., 2005).

These instruments now detect thermal dust emission, atomic and molecular lines (CO, but also HCN, CI and CII) from objects such as quasars, radio galaxies and Lyman break galaxies out to z = 6.4 (see Fig. 10, adapted from Bertoldi et al. (2003) and Maiolino et al., (2005). Blank field surveys with SCUBA and MAMBO (see Blain et al. (2002) for a review) have also revealed a new class of sub-mm galaxies (SMGs). The host galaxies of these SMGs are often very faint at optical/near-infrared wavelengths (eg, Dannerbauer et al., 2004). Since the beamwidths (11" to 15") of the bolometer arrays are large, the identifications have often required deep radio imaging with the VLA (eg, Ivison et al., 2002). Optical spectroscopy of these host galaxies has now determined the redshift of 73 SMGs; the mean is <z> = 2.2 (Chapman et al., 2005). At least 12 of these redshifts have been confirmed with CO spectroscopy (Greve et al., 2005). The importance of CO confirmations is illustrated by the detection of multiple velocity components in several SMGs, and the possibility that in some cases, there may be a foreground galaxy acting as a gravitational lens. The main difficulties in this work have been the low angular resolution of the bolometer arrays and the limited sensitivity and bandwidth of millimetre interferometers.

The objects found so far are the most luminous and thus are very probably non-representative. Several of these are lensed sources. Future measurements will provide a more representative sample of detections. Herschel SPIRE and PACS will be much more sensitive than the ground-based sub-mm bolometers, so deep, wide-field sky surveys will detect thousands of star-forming galaxies for wavelengths near the peak of the thermal dust emission at the median redshift. However, the spatial resolution of SPIRE and PACS is comparable to that of the ground-based bolometer arrays, so the identification and redshift determination of these galaxies will need to rely on other instruments. ALMA will be the main follow-up instrument, since ALMA's high sensitivity and receivers can cover 8 GHz wide slices of the frequency band. This will allow to completely bypass identifications that were formerly made in the radio and optical/near-infrared, and directly measure redshifts using the CO lines (see Fig. 11). Simultaneous high spatial resolution continuum imaging will also allow one to study the morphological distribution of the thermal dust emission, pinpointing the regions of star formation in these galaxies. Very sensitive ALMA observations of these PACS/SPIRE sources will also allow us to search for fainter molecular and atomic emission lines such as the CI transitions at 609 μm and 370 μm. Additional possibilities are the $C^+$ line at 158 μm, the OI lines at 146 μm and 63 μm, NII lines at 203 μm and 122 μm, and OIII lines at 88 μm and 53 μm (see Fig. 1). The detection of the redshifted $C^+$ line in J114816.64+525150.3 (see Fig. 10) shows that ALMA should be able to measure ionised carbon in many high redshift sources (Maiolino et al., 2005). The combined analysis of these molecular and atomic lines using radiative transfer models should allow a determination of the physical parameters of the interstellar medium (ISM) such as temperature, density and chemical composition in these high redshift objects.

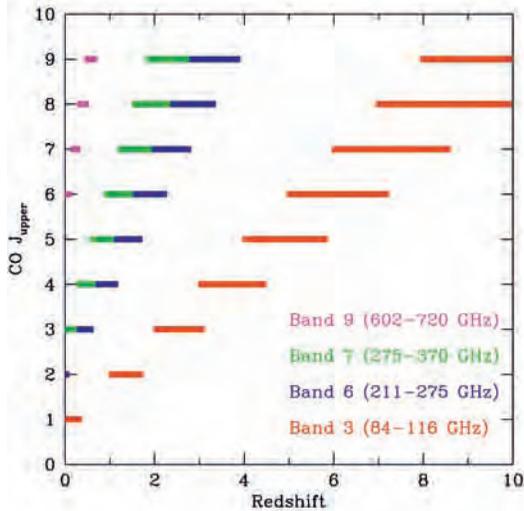

**Figure 11:** This plot shows the coverage of redshift versus the rotational quantum number J for carbon monoxide transitions for the receivers delivered in the North America ESO agreement. This illustrates the possibility of measuring the redshift of molecular content of a source detected only in dust continuum. The colour coding identifies the ALMA "first light" receiver bands (provided by C. de Breuck).

In one scenario PACS and SPIRE measurements provide sources for ALMA follow-up studies. An alternative is described in the DRSP below, in which ALMA is used to make a very deep unbiased survey.

> In DRSP[1] contribution 1.1.1, *Unbiased survey of sub-mm galaxies*, Part 1, it is proposed to carry out an unbiased broadband continuum survey of a 4' by 4' region at 290 GHz. The continuum noise limit is 24 μJy. At 290 GHz, each ALMA antenna has a full width to half power (FWHP) of 21". To fully sample the region, one must space pointing positions by a FWHP. This requires 132 pointings. To reach the rms noise, one must integrate at each position for 30 minutes. The total time is 80 hours; this time includes the measurement of reference positions and calibrations. One could increase the value of such a survey by choosing a region such as the Hubble Deep Field South, or a GOODS region.
>
> A follow-up contribution is a combined spectral line and continuum survey, in DRSP programme 1.1.2 *Unbiased survey of sub-mm galaxies, Part 2*. This is to be carried out at 3 mm, with an angular resolution of 3". The FWHP beam of each antenna is 1', so continuum measurements of a 4' by 4' region require 16 pointings. The time needed is 4 hours per position to reach an RMS noise limit of 1.5 μJy. The spectral line survey is to be carried out using 4 separate settings. Each setting is spaced by 8 GHz, the velocity resolution is 50 kms$^{-1}$, or 16.65 MHz at 100 GHz. To reach the rms noise limit of 40 μJy, one must integrate 4 hours at each position, for a total time of 64 hours.

---

1      The boxed text here and in the following sections are excerpts from the ALMA Design Reference Science Plan describing proposed science applications for ALMA.

In addition to emission lines from the sources, ALMA might detect absorption lines against background sources, even out to high redshifts. Such measurements allow estimates of the densities and temperatures of the line-of-sight gas, and in addition isotopic ratios from measurement of $^{13}$C, $^{15}$N, $^{18}$O and D substitutions. To date molecular hydrogen has been detected in nine systems, but CO has been found in only four absorption systems, of which two are truly intervening and not associated with the background source (Curran et al., 2004). With ALMA it would be possible to detect more molecular species in already known systems as well as finding many new absorption line systems. With the detection of new absorption line systems, one could determine accurate abundances of molecules over cosmic time. Such measurements might be more difficult with Herschel due to baseline uncertainties.

> In DRSP contribution 1.3.1 *Spectral line survey in high-z molecular absorption systems*, it is proposed to measure Bands 3, 6, 7 and 9 completely toward 2 bright compact continuum sources and Centaurus A. This would require 43 hours of time.

## 6.4 Active Galactic Nuclei

These sources show intense MIR and FIR emission in the continuum and also show strong atomic and molecular line emission. MIR and FIR emission may originate in the central region where it is likely that an obscuring torus is present and where star formation takes place in the host galaxy. High resolution imaging is then fundamental to test models of the obscuring torus. The emission of this latter is anisotropic and depends on its geometry and size. High resolution imaging is needed to (a) disentangle star-forming regions and the nuclear torus, (b) test the formation model of AGN and (c) investigate the joint formation of AGN and host galaxies. AGNs could be divided into two subclasses according to the rate of black hole growth: one would be a rapidly growing phase and the other a slowly growing phase. Narrow line and broad line Seyfert 1 galaxies are possible candidates for these distinct phases. Joint formation models predict that an efficient feedback provided by star formation is at work in massive systems and halts black hole growth and star formation. The mechanism does not work in less massive and smaller objects.

An understanding of AGN physics and formation, which probably includes a phase involving dust obscuration and black hole growth, can be studied only by carrying out Herschel surveys at 75 μm. Such surveys will detect the dust emission peak of AGN with redshifts of around z = 0.7 to 1. These are the objects responsible for the hard X-ray background and are crucial for a determination of the AGN bolometric luminosity, accretion rate and black hole mass. At present the accretion rate is estimated from optical and X-ray observations using very uncertain bolometric corrections.

In summary, Herschel/PACS data will define the AGN spectral energy distributions during the dust enshrouded phase, as well as discovering obscured AGNs (so-called Type II AGNs). The FIR SED of Type I objects will also be measured. ALMA will study the details of the nuclear and disk emission using high angular resolution images. With such data, one can test scenarios for AGN evolution, models for the formation of black holes and the concept of the joint formation of AGNs and their host galaxies. With a large data collection one can classify sources, and try to model the AGN as a function of other properties of the source.

> In DRSP contribution 1.51, it is proposed to image six AGNs at distances of 17 Mpc with an angular resolution of 0.06" in the CO J = 2-1 line. The goal is to determine the molecular gas content close to the black hole at the centre of the AGN, to estimate the mass of infalling material. This angular resolution is equivalent to 5 pc at 17 Mpc. There would be a single pointing, with 5 kms$^{-1}$ velocity resolution. The rms noise is 0.2 mJy or 1.2 K in each beam. This allows the detection of warm CO, and in addition the continuum sensitivity is 4mJy. This limit corresponds to an H$_2$ column density of 1.5 x 10$^{23}$ cm$^{-2}$ if the dust temperature is 20 K, and if the dust properties and gas-to-dust ratios are standard.

## 6.5    Normal Spirals and Low Surface Brightness Galaxies

The shape of the SED from the FIR to sub-mm clarifies the processes at work in distant target objects. There are three different effects: the existence of an active nucleus, the star formation activity of the galaxy and metallicity. Our ability to distinguish between these in a high redshift object must rely on the amount of knowledge we have on a nearby well understood sample. In both photometric and spectrometric modes, Herschel will map several samples of local objects of different type, morphology and metallicity. These data will be used to explore a large range of galaxy properties: the processes controlling the dust physics (size, distribution, composition and evolution), the structure of the ISM, the metallicity, the consequences of heating and cooling on dust in different metallicity environments, how much star formation activity is dust enshrouded and the optical depth of dust to the NIR/MIR, the relation between metals in the gas phase and those in dust as a function of metallicity.

However Herschel data alone will not be able to characterise these processes in detail. Only high resolution measurements with ALMA will probe the detailed physical processes in the ISM under a variety of conditions.

Even in very extended nearby galaxies such as M31 only a few regions have been imaged with millimetre interferometers (see the review by C. Wilson, 2005). The full 2º extent of M31 has only been mapped with single dish instruments with a typical resolution of ~23" in the CO(1-0) line (Nieten et al., 2006). Interferometer maps of nearby galaxies have mostly focused on Local Group members, which are close enough to resolve individual Giant Molecular Clouds (GMCs). In the Milky Way Galaxy, GMCs represent the main sites of star formation. Therefore, studying GMCs in external galaxies of different morphological types and various star formation rates, will help to understand the efficiency of star formation in different environments. For example, the mass function of GMCs appears much steeper in M33 than it is in the Milky Way, suggesting a different organisation of the molecular ISM. With ALMA, it will be possible to resolve GMCs and study their size and turbulence in galaxies out to distances of 100 Mpc.

The study of interacting galaxies will improve our knowledge of star formation triggered by interactions. Consider the nearby M81 group at a distance of 3.8 Mpc (eg, Walter et al., 2005). This is a well-known group of interacting galaxies and has some 40 members representing all morphological types. The most well-known members of this group are the spiral M81, the starburst M82 and dwarf starburst NGC 3077. These have been studied extensively in the HI 21 cm line. Tidal arms and tails stretching out over more that 70 kpc connect the three galaxies. Tidal effects play a large role in their star formation rates and there appears to be a variety of triggered star formation in this group. For M82, there is intense activity apparently leading to the large number of supernova remnants (SNRs). Near NGC 3077, there is an extended tidal arm that contains HII regions over a wide area. Star formation may have just begun in this arm. In systems such as the M81 group, one can use ALMA to measure the CO abundance, and its kinematics, to obtain a description of the gas mass and dynamics. With Herschel it will be possible to determine the dust masses and temperatures, as well as trace atomic lines. With these results one can compare the dust and gas properties and relate these to the local star formation as a function of location within each galaxy. From linewidths and projected positions, we can estimate what portion of the star formation is caused or triggered by tidal forces. It would be interesting to compare interacting group members with other members of the group that do not seem to be affected by tidal forces.

Low Surface Brightness (LSB) and Dwarf Galaxies are among the most common spiral galaxies. These have high neutral gas fractions and are metal poor (eg, Bothun et al., 1997). Despite their low gas densities, these show signs of low-level, ongoing star formation, as indicated by their blue colours and H$\alpha$ emission. The suppressed star formation must result from different physical conditions in the ISM.

To date, dust emission spectra have been observed in a handful of metal-poor galaxies in the local Universe. These galaxies appear to differ remarkably from those with normal metallicity starbursts, emphasising the need to explore the metallicity dependence of SEDs. Dust size distributions seem to be dominated by very small dust grains. In particular in the sub-mm/mm range, an excess of very cold dust has been observed. Dust and gas properties, heating and cooling processes will undoubtedly be altered relative to dust-rich galaxies.

PACS and SPIRE photometry and spectrometry of dwarf galaxies at different metallicities can be used to construct the emission spectrum of the dust and gas that have not yet experienced repeated recycling through the ISM.

Molecular line surveys will help to constrain the chemistry and may help to determine whether molecules are located mostly in Giant Molecular Clouds (GMCs) or in a more diffuse medium (eg, Matthews et al., 2004). The molecular abundances are expected to be low, so the measurements of both Herschel and ALMA may be restricted to objects with redshifts of z~0.5 or less.

If very nearby, such sources are usually more extended. Thus the beam switched data from PACS/SPIRE measurements will require corrections to determine the total flux densities if the source is larger than the chopper throw. In the case of ALMA, one can use single dish measurements to provide the total source flux density. If the angular sizes of these sources may exceed the primary beam size of the ALMA antennas, mosaics will be required to image the entire source. For ALMA this can require long integration times. As an example, in Fig. 12 we show the dwarf galaxy IC10, together with the Herschel and ALMA beams. IC10 is a northern hemisphere source. More southerly analogues are the Large and Small Magellanic Clouds. These nearby dwarf irregular galaxies have been the subject of a key programme at the SEST radio telescope. With ALMA, one can even detect continuum emission from individual asymptotic giant branch stars in the Magellanic Clouds.

> In DRSP programme 1.7.1, it is planned to carry out a line survey toward the inner regions of three types of galaxies, one of which is an M83 type spiral galaxy. For this source, it is planned to use 4 frequency settings to cover 32 GHz. The frequencies would be in the range 218 to 270 GHz. The angular resolution is to be 1", with measurements made at 6 positions. For integration times of 2 hours, the rms noise in a 5 kms$^{-1}$ wide channel is 10 mK. It is estimated that that more than 100 transitions could be detected, and the results could be used to carry out modelling of the chemistry of the galaxies, including the role of gas phase and dust chemistry. For 6 positions in 3 sources, with 4 velocity settings, the total time is 36 hours.
>
> For dwarf galaxies, in DRSP contribution 1.7.2, it is planned to measure 10 low surface brightness galaxies in the J = 1-0 line of CO at 3 mm and follow-up measurements of the in the J = 3-2 line of $^{13}$CO at 1 mm. The galaxies will be at 10 to 20 Mpc, so that the CO lines will be redshifted by 700 to 1400 kms$^{-1}$. The angular resolution at 3 mm will be 1", corresponding to 50 to 100 pc. The velocity resolution will be 5 kms$^{-1}$. With these parameters, the sensitivity is 0.02 K in the CO J = 1-0 line after 5 hours. For the $^{13}$CO line the sensitivity is 6 mK.

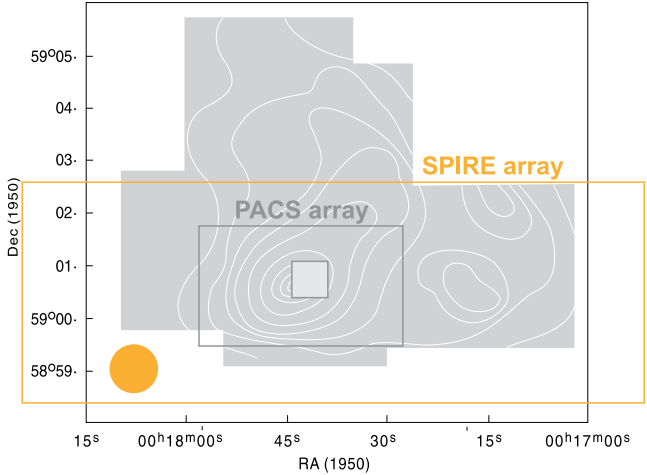

**Figure 12:** Data for IC10. The integrated CII emission is shown as contours. The shaded region is the sampled area (Madden et al., 1997). The rectangles show the field of view of the SPIRE array and the PACS array of Herschel. The smallest rectangle shows the size of the spectrometer.

## 6.6 Star Formation and Molecular Clouds

### 6.6.1 Protostars

The study of the star formation process is one of the primary science goals of ALMA. In Fig. 13 we show a sketch of the four stages of star formation, from the collapse of a molecular cloud to the formation of a star surrounded by a disk. For column densities, $N > 10^{18}$ cm$^{-2}$ and densities $n > 10^2$ cm$^{-3}$, interstellar gas consists of molecular hydrogen, $H_2$, and helium. The $H_2$ molecule will not produce emission lines if kinetic temperatures are below ~100K and there are no shock waves. Then the abundances of the $H_2$ molecules must be traced indirectly. Cloud collapse requires high interstellar gas densities and low kinetic temperatures. Under such conditions, grain properties change and the constituents of the gas will condense onto grains. With the broadband data from SPIRE and PACS one can accurately determine the total luminosity and with the spectrometers, measure fine structure lines of atomic species. With Herschel HIFI it will be possible to take spectra without absorption in the Earth's atmosphere. This is especially useful for $H_2O$. The higher angular resolution ALMA images will help to refine the analysis of models based on Herschel data. The final result will be the distribution of $H_2$ (which has no spectral lines at low temperatures), of selected atoms, a number of molecules and dust, as well as the dynamics.

Herschel HIFI can measure spectra with a velocity resolution of better than 0.1 kms$^{-1}$, while ALMA can measure spectra with a velocity resolution exceeding 0.01 kms$^{-1}$. Measurements with such a fine velocity resolution are important since models show that the collapse velocities of low mass protostars are of order 0.1 kms$^{-1}$.

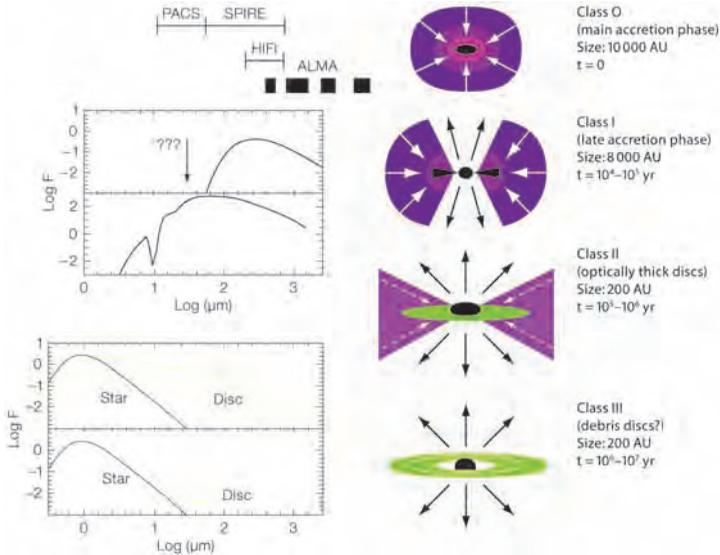

**Figure 13:** Right side: A sketch of the development of a low mass protostar and its disk. The wavelength coverage of the Herschel instruments PACS, SPIRE and Herschel HIFI are shown above left. The ALMA receiver bands from left to right are Band 9, Bands 7 and 6 and Band 3 in the bilateral ALMA project. With the addition of Band 5 and Bands 4, 8 and 10, the coverage of ALMA receiver bands provide a solid block in the uppermost part of the figure under "ALMA". These will also fill the longer wavelength part of Herschel coverage (after Shu et al., 1987, Lada, 1987 and unpublished plots by M. Hogerheijde).

As can be seen from Fig. 13, ALMA can be used to study star formation at longer wavelengths, whereas PACS and SPIRE can be used to determine the SED shortward of the peak of the luminosity curve.

The development of protostars in the Class II and III phases involves outflows. Until recently these have been traced using CO lines. However this method is dependent on geometry. An alternative would be the use of the OI line, at 63 μm, measured with PACS (eg, Walmsley & van der Tak, 2005). With SPIRE, one could measure the dust emission at longer wavelengths to obtain total masses. With ALMA, higher resolution images of dust and gas phase constituents would give a determination of the density and kinetic temperature distributions. Many species may have condensed onto dust grains in protostars, but there has been evidence that species such as $H_2D^+$ are less condensed at very high densities. High angular resolution spectral line measurements are needed to fully specify the chemistry (eg, Herbst, 2005; Flower et al., 2005).

Water vapour is one of the most abundant and important molecules in star-forming regions. In cold, dense molecular clouds, water is only a trace species. However, in warm regions close to newly-formed stars, water vapour becomes the third most abundant species, after the usually unobservable $H_2$ and helium. These regions are the inner protostellar envelopes where the dust is warmer than the ice evaporation temperature, and where the collapsing matter interacts with the powerful jets from the protostar causing violent shocks (Olofsson et al., 2003; Boonman et al., 2003). This enormous variation in water vapour abundance makes it a unique probe of the physical structure of the region and of the fundamental chemical processes in the gas and between the gas and the grains. Moreover, deuterated water provides an important record of the temperature history of the object and the conditions during grain surface formation. Water vapour also plays an active role in the energy balance. Because it has a very large dipole moment, water vapour emission lines can be efficient coolants of the gas. All these aspects make water vapour an important probe of star-forming regions: $H_2O$ provides highly complementary information to that derived from the commonly studied CO molecule.

Except for a few maser lines, measurements of thermal emission of non-redshifted water vapour are not possible due to absorption in the Earth's atmosphere. Thus, the ability to measure $H_2O$ with Herschel HIFI is particularly important. In Fig. 14 we show the transitions of water vapour that can be measured with Herschel HIFI. The finest angular resolution of Herschel HIFI is 13". Additional water vapour transitions fall within the passbands of the PACS and SPIRE spectrometers. These lines can be measured with angular resolutions to 9", but these lines will not be velocity resolved. Thus ALMA images of protostellar regions with a sub-arc second resolution are needed to supplement the Herschel HIFI data. For a specific source, spectral line surveys made with Herschel HIFI could be supplemented by higher resolution ALMA images of the $3_{13}$-$2_{20}$ lines of $H_2O$ and $H_2^{18}O$ (using ALMA Band 5 receivers) and other species. ALMA measurements allow a better estimate of source sizes and thus the true abundances of species rather than the beam-averaged column densities. Such source averaged column densities are those needed for chemistry models. For some sources, the angular resolution of ALMA is needed to determine angular sizes. Herschel would provide the total abundances of a species such as $H_2O$ integrated over the source. The abundance of $H_2O$ in each part of a source would depend on ALMA images. This will have an important influence on the energy balance and chemistry. These considerations apply to both active high mass star-forming regions, molecular clouds, envelopes of AGB stars and solar system objects.

As species condense onto grains or grains coagulate, the spectral indices will change. Measurements of dust emission may allow a determination of the extent of these processes. PACS and SPIRE will provide such data, supplemented by ALMA images; see Su et al. (2005) for a sample of Spitzer results in this area.

> In DRSP contribution 2.3.1 *Chemical survey of hot cores and inner warm envelopes around YSO*, it is planned to image a large sample of protostars in a variety of molecules that probe different chemical and physical processes, for example, ice evapouration, hot gas chemistry leading to complex organics, shocks, infall, and the formation of disks. The continuum data combined with radiative transfer modelling provide constraints on the physical structure (temperature, density profile) of the region, whereas the line data constrain the abundances as a function of position in the source. Multiline/multiband observations are needed to separate excitation effects from abundance effects. The spatial resolution is 0.2" and the rms noise in a 0.25 kms$^{-1}$ channel is 1 K for a typical 1 hr integration per frequency setting. About 10 frequency settings in Bands 6, 7 and 9 are proposed, with each setting expected to contain dozens of detectable lines. Both single dish and ACA data are needed to obtain accurate images. For 65 sources, this programme takes 585 hr.

The above ALMA programme could be coordinated with Herschel HIFI and PACS surveys of a large range of water lines and related hydrides (OH, $H_3O_+$) in the same objects. A combined ALMA-Herschel programme would allow measurements of all the major ice and gas components.

> A combined ALMA-Herschel programme would address all the major ice and gas components. In DRSP contribution, 2.1.4, *Density and Temperature Profile in High Mass Cores*, it is planned to measure the properties of sources similar to that in the uppermost panel of Fig. 10. The angular resolution is to be 0.5". The contribution includes measurements of 6 transitions of $H_2CO$ at 0.8 mm to determine kinetic temperatures $T_{kin}$, as well as measurements of broadband dust emission. The maximum line temperature is expected to be 7 K. The rms noise in a 0.5 kms$^{-1}$ spectral channel is 0.15 Kelvin. The rms noise in a broadband continuum measurement is 30 mJy beam$^{-1}$. When combined with the $T_{kin}$ values, a dust absorption coefficient and a dust-to-gas ratio, measurements of the broadband dust emission allow a determination of $H_2$ column densities and, with assumptions about geometry, local densities. For a dust temperature of 10 K, the broadband continuum noise gives a limit to the $H_2$ column density of $10^{22}$ cm$^{-2}$. The sources are 1 to 10 kpc from the Sun. For one half of the sources, one requires a 5 position by 5 position mosaic with 1.5 hours integration per field. For the more distant sources, a 3 position by 3 position mosaic with one hour of integration per field is needed. The total ALMA observing time is 155 hours. For accurate images, one requires both single dish flux densities and also measurements with the ACA. We take a total velocity range of 10 kms$^{-1}$, and a total image size of 100" by 100".
>
> Without the ACA or single dish data, each of these datasets has more than 6 x 10$^7$ values for the spectral lines alone. Such a data set cannot be compared to models without elaborate computer programmes which optimise the fit by extensively (and intelligently) varying model parameters. The initial guesses for geometries must come from theory, and previous measurements in the infrared.

In DRSP contribution 2.1.2, *Kinetics, Density and Temperature Profile in Pre-Stellar Cores*, it is planned to measure the *J = 2-1* transitions in carbon monoxide, CO, and isotopes, and the *J = 1-0* transition of $H_2D^+$, as well as broadband continuum. Measurements of the $H_2D^+$ molecule are interesting since this appears not to deplete onto grains at high $H_2$ densities. The sources are 160 pc from the Sun, so the 1" angular resolution corresponds to $2.3 \times 10^{15}$ cm, or 160 Astronomical Units. In addition to interferometer data, results from the ACA and single telescopes are required. The peak line temperatures will be ~3 K for $C^{18}O$, 5 K for $^{13}CO$ and 0.1 to 1 K for $H_2D^+$. For the optically thin line of $C^{18}O$, over a wide range of densities and kinetic temperatures, the rms noise in a spectral channel gives a limit to the $H_2$ column density of $4 \times 10^{21}$ cm$^{-2}$. The rms noise in continuum measurements is 5 mJy beam$^{-1}$. This RMS noise gives an $H_2$ column density of $5 \times 10^{20}$ cm$^{-2}$ if the dust temperature is 10 K. Since the *J = 1-0* transition of $H_2D^+$, is absorbed in the atmosphere of the Earth, 100 hours are needed for these measurements, while 25 hours are needed for measurements of the CO transitions.

For both DRSP programmes, the absolute calibration accuracy should be better than 10% for continuum and spectral line measurements. The repeatability should be 3% and the relative calibration between receiver bands 3%.

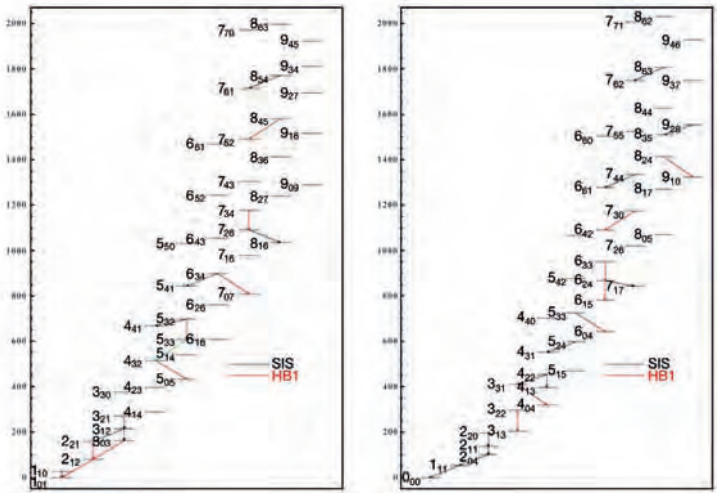

**Figure 14:** The lower lying energy levels of the water molecule. On the left are the levels for ortho-$H_2O$, on the right for para-$H_2O$. The energy levels connected by vertical lines are the transitions to be measured with Herschel HIFI. A few transitions have been measured from the ground. The strong maser line at 22 GHz from ortho-$H_2O$ is the most widely observed. The 183 GHz line of para-$H_2O$ will be in Band 5 of ALMA (provided by J. Cernicharo).

## 6.7 Selected Galactic Sources

There are a number of galactic sources that are extensively studied because of their chemical richness, dynamics or stage of evolution. In all of these sources, water vapour is an important constituent, so there will be surveys of water vapour line with Herschel HIFI. In many cases these sources have a range of physical conditions, so ALMA images in both molecular, CI and dust continuum will be needed to interpret the Herschel data. Spectral line surveys can also be carried out with ALMA. These have the advantage of high angular resolution so that the different regions can be spatially resolved (eg, van Dishoeck et al., 2005).

### 6.7.1 Spectral Line Measurements

The Sagittarius B2 region is at a projected distance of 104 pc from Sgr A*, the centre of our galaxy (for an assumed distance of the Sun to the galactic center of 8.5 kpc). Sgr B2 is one of the most active star-forming regions in the galaxy (eg, Nummelin et al., 2000). It is also a region where complex molecules are often found. The structure of the Sgr B2 cloud is complex, consisting of a very high density compact core within an extended low density envelope. There are measurements of thermal water vapour lines toward Sgr B2 from satellites such as SWAS or ODIN, ISO or the Kuiper Airborne Observatory (eg, Cernicharo & Crovisier, 2005). These show a region rich in water vapour, but with large differences in density and temperature along the line of sight (see Comito et al., 2003) for a model of $H_2O$ emission). The radial velocities of the core and envelope are similar, so an extension of water vapour measurements with Herschel HIFI with spectral line and dust continuum images made with ALMA would allow further refinements leading to a more complete understanding of conditions in Sgr B2 and the galactic centre region.

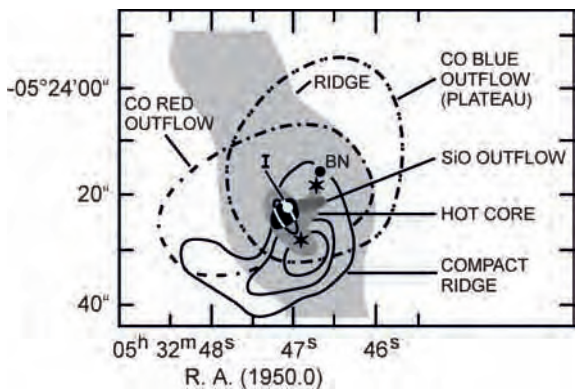

**Figure 15:** A sketch of the region close to the Kleinmann-Low nebula is shown. The shaded region in the NE to SW direction is the Ridge component, which is a part of very extended emission. The hot core is a warm dense region where a wide variety of complex molecules are found. The compact ridge is well defined by species such as methanol, whereas the hot core shows a large abundance of nitrogen bearing species. The SiO and CO bipolar outflows are thought to be caused by source "I", which is very likely to be an extremely young B star (Wilson, unpublished).

The Orion KL region is one of the nearest of the active regions of massive star formation, about 500 pc from the Sun. The molecular cloud is behind the HII region, making optical measurements difficult. In Fig. 15 we show a sketch of the Orion KL region and in Fig. 16, a sub-mm spectrum from this source. The KL nebula in the molecular cloud is a < 30" size region where O and B stars are now forming. Thus it might be possible to study the earliest phases of high mass star formation. The KL nebula contains a "hot core" region which has $H_2$ densities of order $10^7$ cm$^{-3}$, and temperatures of order 150 K. The hot core contains a variety of complex molecules and outflows, providing a view of complex interstellar chemistry.

The DRSP contribution 2.3.4 *Unbiased Line Surveys of High Mass Star-Forming Regions* is an ambitious programme to survey Orion KL and 9 other galactic sources in all four ALMA receiver bands. For the 3 mm band this requires 60 frequency settings, with integrations of 10 minutes at each setting. With a velocity resolution of 0.1 kms$^{-1}$ this programme would require 100 hours for the measurements. In DRSP contribution 2.3.5, *Low Frequency Spectral Survey Aimed at Complex Organics*, proposes a spectral line survey of Sgr B2 and Orion, NGC 6334, Leo and Aquila. Since this uses a velocity resolution of 0.3 kms$^{-1}$ the integration time is commensurately smaller.

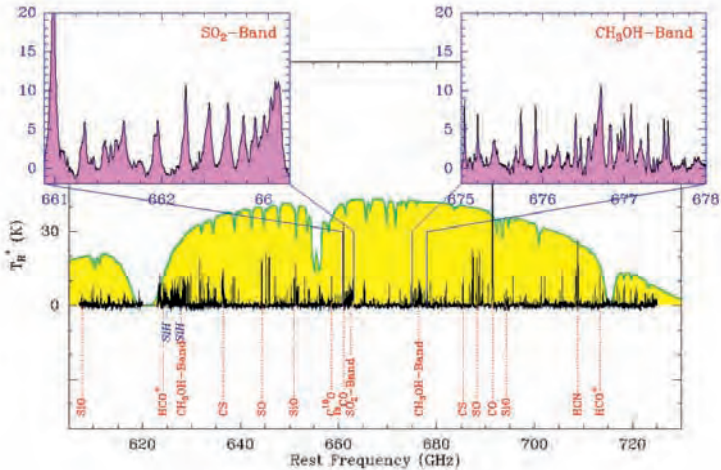

**Figure 16:** The combined spectrum from a survey carried out by Schilke et al. (2001) with the Caltech Submillimeter telescope on Mauna Kea. The angular resolution is 13" to 11" over this frequency range. The temperature scale has been corrected for absorption by the atmosphere. The plot in yellow shows the transmission through the atmosphere. A few of the more intense spectral lines are identified. Emission bands from the sulphur dioxide and methanol molecules are shown in two inserts.

There have been a few fully sampled images of extended cold dust clouds (Panis et al., 1998; Falgarone et al., 2005) in a few spectral lines, usually of carbon monoxide and its isotopes. Such images help to characterise sources before stars form. With ALMA such imaging can be carried out in a number of different spectral lines simultaneously.

> In DRSP 2.1.1 *Small Scale Structure of Molecular Clouds*, there is a proposal to image a very extended low bright object such as a molecular cloud without embedded stars. The plan is to image a 4' by 4' region with an angular resolution of 1", a velocity resolution of 0.1 kms$^{-1}$, and a rms noise of 0.3 K in the J = 1-0 lines of CO and $^{13}$CO at 3 mm wavelength. To reach this noise level in one position requires 1 hour in the CO line. If both lines can be measured simultaneously, this project requires 64 hours of measurement time. Because of the multi-beam bolometers, a comparable image in dust continuum is more easily carried out with PACS or SPIRE, but with lower angular resolutions.

### 6.7.2    Continuum Measurements

The Sagittarius A* region is the centre of our galaxy, at a distance of 8.5 kpc from the Sun. Sgr A* is a point source, but is surrounded by a great deal of cool dust emission and thermal free-free emission from a ring-like structure. The inner 1' of the galactic centre is a superposition of these sources, so the result is a very complex region (eg, Morris & Serabyn, 1996). For this reason, measurements with PACS or SPIRE will be difficult to interpret without additional data. To search for short term time variability from the non-thermal Sgr A* emission, one could use a series of ALMA measurements, each made with five minutes of integration. The sensitivity is more than sufficient to detect 1% variations. If the data are taken in the most extended array configuration, the data would discriminate against low brightness thermal emission from dust and Bremstrahhlung, leaving an image of the point source Sgr A*. Spatially resolving Sgr A* is of great interest since this is the only opportunity to resolve a massive black hole. This measurement would involve millimetre/sub-mm Very Long Baseline Interferometry, with the entire ALMA serving as one of the elements of the array (eg, Carilli, 2005)

There have been detections of O and B star emission in Orion KL (Zapata et al., 2004). These measurements show rising spectra between 7 mm and 1.3 cm, so that the detection of high mass stars should be possible in short integration times with ALMA. There might be a serious problem with confusion in the larger Herschel beams, for sources near the galactic plane. However for objects at high galactic latitude, it should be possible to isolate the stars sufficiently to make unambiguous comparisons of Herschel results with ALMA images.

## 6.8    Large Scale Surveys of our Galaxy

Such surveys will give unbiased estimates of the number of different types of sources. Surveys may also lead to the discovery of new types of sources, or better examples of known types of sources. An example of a large scale bolometer image from a satellite is the ISOCAM survey of the galactic plane (Omont et al., 2005). This covered 16 square degrees. The GLIMPSE survey is being carried out with the Spitzer satellite (Uzpen et al., 2005). There is also a plan to map a 100 square degree region with SPIRE and PACS. Clearly, ALMA will be used to follow up such surveys, but it is unlikely that ALMA would be used in a survey mode for mapping large areas in the broadband continuum. There are no comparable programmes in the ALMA DRSP.

## 6.9  Evolved Stars

### 6.9.1  Spectral Lines

In the late stages of stellar evolution, lower mass stars go through the asymptotic giant branch, AGB, on their way to becoming a planetary nebula. The AGB stars have extensive envelopes which contain molecules (eg, Olofsson, 2005). The nearest AGB star with such an envelope is R Leonis, also known as IRC+10216 (see Fig. 17). Such envelopes are close to being spherically symmetric, so are more easily modelled. There have been extensive line surveys of this source, but further examples should be studied to determine molecular, atomic and isotopic abundances. We show an example of a few images of different molecular lines in IRC+10216 in Fig. 17. Clearly the spatial distribution of CS or SiS is very different from CN or $SiC_2$. To determine the chemical development of the envelope, one needs to have the abundance and spatial distribution of many such species.

After the AGB phase, the star passes though a short-lived proto-planetary nebula phase while evolving to a planetary nebula (PN). The proto-PNs have intense UV fields and outflows; two of the most studied are CRL 618 and 2688 (eg, Millar & Woods, 2005). With Herschel, one can obtain accurate beam-averaged abundances of gas phase species and dust, but high resolution images made with ALMA are needed to determine the actual abundances since the source sizes are of order 10" or less.

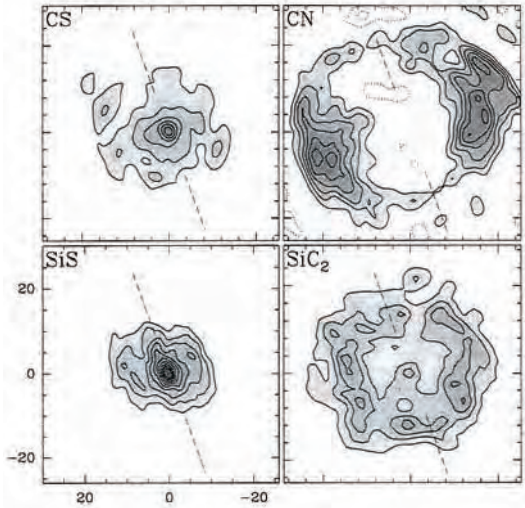

**Figure 17:** Images of emission from IRC+10216 from the species identified in the upper left corner of each panel. The contours are the integrated intensities. IRC+10216 is a solar mass star on the Asymptotic Giant Branch (AGB). Images from Lucas et al., 1995.

The high spatial resolution, large FOV and good sensitivity of Herschel will lead to great progress in our understanding of the mass loss process in evolved stars, the time evolution of the mass loss rate, how it influences the shaping of the surrounding nebula, properties of the dust formed and the formation site of the dust. Spectra of AGB stars and PNe are ideal tracers of solid and molecular species that form in the cool circumstellar envelope. Most of the astronomical solid state features are found in the near and mid-infrared range. However, dust species like Forsterite $Mg_2SiO_4$ at 69 μm (a sensitive dust thermometer), Calicite ($CaCO_3$) at 92.6 μm, crystalline water-ice at 61 μm, and other dust species and ices will be measured by Herschel in our galaxy and other nearby galaxies. In addition in some circumstellar envelopes, powerful $H_2O$ and SiO masers are present. These will be easily detected with ALMA. ALMA can image the SiO quasi-thermal millimetre line emission in the outflows of the stars. With distance from the star, the SiO molecule first freezes out onto dust grains as the temperature drops; further still, SiO is photodissociated by the interstellar UV-radiation field. Measurements of SiO, when combined with CO line emission will be used to follow the envelope evolution (gas expansion and acceleration) within 20-30 stellar radii. These data will provide a database of dust formation and evolved-star physics that will be used as a template to study the spectra of high-redshift galaxies that will be targeted with ALMA.

> An example is DRSP contribution 3.3.2, *Line Surveys in Evolved Stars*. The plan is to produce images like those in Fig 17 for 6 objects other than IRC+10216 and also for higher frequencies. The survey will include 6 sources, 4 ALMA receiver bands. It will use an angular resolution of 0.6" to 0.3" and a velocity resolution of 0.5 kms$^{-1}$. The rms noise in the spectral lines will be better than 1 Kelvin in 10 minutes of integration time. The continuum rms will be better than 0.01 Jy. The programme will require 170 hours of time.

## 6.9.2 Continuum Emission

Another research topic is the search for continuum radio emission from red giant and supergiant stars. From 1.3 cm and 7 mm measurements, Reid & Menten (1997) found that such emission has an optically thick thermal spectrum.

> In DRSP 3.2.3, *Thermal Emission from Red Giant and Supergiant Stars*, it is proposed to measure a number of spectral lines in the envelopes of well-known stars such as o Ceti, R Leonis, W Hydrae, R Aquilae, Chi Cygni and R Cassiopeiae. It is estimated that the continuum emission from these nearby objects can be detected with ALMA in 10 seconds. In 30 minutes integration, one reaches 7 μJy, so that one can detect G and M dwarf stars at 10 pc distance at 100 GHz. In DRSP 3.2.6, *Millimetre Survey of Stellar Disk Emission from late-type Stars*, R. Osten proposes to detect the emission from 20 stars in 3.5 hours. A repeat of these measurements would be made more than 1 month late to search for variability. In any such programme, the highest angular resolution is needed to eliminate confusion. In DRSP 3.2.7, *Flares from Young Stellar Objects: what we learned from the 2003 January flare in GMR-A* it is proposed to integrate for a short time on a large number of objects in the three lowest frequency ALMA receiver bands to search for variability caused by magnetic surface activity.

## 6.9.3 Solar System Objects

Studies of the outer planets are often searching for the origin of external water. The sources might be dust from asteroids or comets, or local production. These sources can be distinguished by measuring Deuterium/Hydrogen ratios. At present, there seems to be different D/H ratios for comets and the outer planets, including Titan. Further studies might confirm this result and lead to more secure models. Herschel would concentrate on measurements of H$_2$O, while ALMA would be used to determine HDO abundances; by combining these data one would obtain reliable D/H ratios. An associated study would be the relation of water vapour production and evolution in comets as a function of heliocentric distance.

For Herschel, water vapour excitation would be an extensive study in comets, and in additional searches for small water abundances could be carried out for asteroids, distant comets and Centaur objects. This would include short period comets formed in the Kuiper Belt, and the measurements should be extended to deuterated water to determine the D/H ratios. These may require simultaneous measurements.

Mars is another prime object of study. One could use Herschel for a spectral line survey in the 960 to 1250 GHz range, and then use ALMA for follow-up studies and confirmations.

> DRSP 4.1.3, *Chemistry in the atmospheres of Venus and Mars*, is a proposal to search for and then image mm/sub-mm spectral lines of molecules such as $SO_2$, $H_2S$, HCl, $^{16}O^{18}O$, $H_2CO$, $H_2O_2$ and NO. This will require 92 hours, and should be repeated two years later for the purpose of monitoring. There is an additional DRSP, 4.1.2, to study HDO from Mars. In DRSP 4.1.4, it is proposed to study the stratospheres of the giant planets in lines of CO, HCN, CS, $HC_3N$ and HDO. There would also be blind searches; in all, this would require 154 hours. In DRSP contribution 4.1.8, it is proposed to measure the atmospheres of Triton and Pluto in spectral lines, and DRSP 4.2.1 proposes to study the albedos sizes and surface properties of transneptunian objects.

### 6.9.4  Modelling and Analysis of the Data

The data sets that will be produced by ALMA and Herschel will be so large that the analysis and comparisons with models will have to be streamlined and automated as much as possible. Various groups around the world have started to address this issue and substantial progress is being made.

For molecular clouds, chemical abundances and physical parameters have been derived traditionally using simple tools, such as rotation diagrams assuming a constant excitation temperature $T_{ex}$ or statistical equilibrium calculations using an escape probability approximation model for a single temperature, T, and density, n. In both cases, comparison with a "standard" molecule such as CO is needed to derive the abundance with respect to $H_2$. It is well-known that this method can lead to abundances that are in error by more than an order of magnitude. For example, the use of $n = 10^5$ versus $10^6$ cm$^{-3}$ for the interpretation of the HCN J = 4-3 line with a critical density, n*, of ~$10^7$ cm$^{-3}$ gives a factor of 10 difference in derived HCN abundance. Densities are known to range from $10^4$ to >$10^7$ cm$^{-3}$ across star-forming regions. Also the neglect of beam dilution and optical depth effects can result in large errors. For example, the emission from the complex organic molecules may originate in a region of only 1", much smaller than the observational single-dish beam of >15" which applies to the more extended CO.

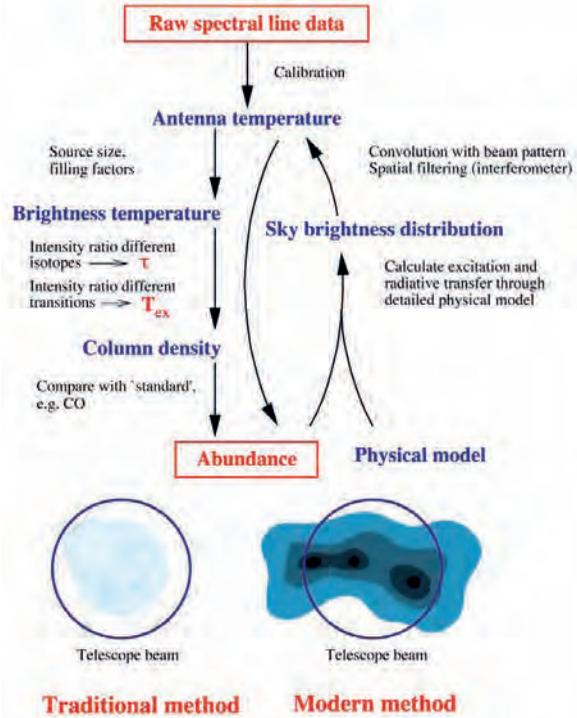

**Figure 18:** A schematic of the analysis needed to convert temperatures into physical models of molecular clouds (from van Dishoeck and Hogerheijde, 1999).The two sketches at the bottom indicate the difference between the use of highly symmetric, unrealistic models (on the left) and more realistic representations obtained from large data sets by the use of an automated modelling process (on the right).

Modern analysis methods start with a physical model of the source, constrained by observations. The excitation and radiative transfer are subsequently calculated through this model and the resulting sky brightness distribution is convolved with the beam profile, or, in the case of interferometer data, analysed with the same spatial filtering. The abundance of the molecule is then adjusted until agreement with the observational data is reached, eg, through a $\chi^2$-test (see Fig. 18; also, eg, Shirley et al., 2002; Jørgensen et al., 2002; Doty et al., 2004, for a summary of the modelling procedures). Basic molecular data such as Einstein A coefficients and collision rates form essential inputs (for a summary, see Schöier et al., 2002, and http://www.strw.leidenuninv.nl/~moldata).

Such an approach can be extended to the study of extragalactic sources. However, the measurements will be less sensitive and within each resolution element there will be a number of different objects. Thus the models require further inputs to determine definite results. For planets, such detailed models have been used for some time. With the additional imaging data from ALMA, in addition to the contributions from Herschel, the size of databases and sophistication of models will increase dramatically.

# 7    Summary

The PACS and SPIRE bolometer systems on Herschel are well suited to surveying large regions of the sky, whereas ALMA can provide high sensitivity, high angular resolution images in spectral line and continuum at longer wavelengths for regions of a few arc minutes in size. Thus, ALMA is better suited to be a follow-up instrument for Herschel surveys. Such follow-ups could be measurements of CO rotational transitions to determine the redshifts of sources detected in the dust continuum or broadband continuum to provide the component of Spectral Energy Distributions (SEDs).

For spectral lines, ALMA will be complementary to Herschel because of different angular resolutions, frequency ranges and effect of the Earth's atmosphere. For any spectral line surveys with Herschel, follow-up measurements with ALMA will greatly increase the scientific value, but it must be stressed that there should be an effort to make the Herschel surveys as complete as possible. ALMA images of many spectral lines will allow better estimates of source sizes. With Herschel one can measure line intensities for entire sources, but ALMA images would give estimates of the true source sizes, and thus the actual abundances of species, exactly those needed for chemistry models. Water vapour is a case of particular interest. High angular resolution images of the 183 GHz line of $H_2O$ and the 203 GHz line of $H_2^{18}O$ in ALMA Band 5 will be useful to supplement the lower angular resolution measurements of many lines of water vapour planned with Herschel HIFI. For protostars and molecular clouds the abundance of $H_2O$ will have an important influence on the energy balance and chemistry. In the study of protostars, the angular resolution of ALMA is needed to determine the sizes of components and variations of abundances on scales finer than a few arc seconds. Herschel would provide measurements of SEDs and line transitions that cannot be done with ALMA. Similar considerations apply to envelopes of AGB stars and Solar System objects.

# 8 Recommendations for the Future

A number of conditions must be fulfilled to combine Herschel and ALMA datasets. First, the sources must be observable with both instruments. For ALMA this condition requires that the sources rise to more than 30° elevation at the ALMA site, with higher elevations even better. Second, the calibrations for both instruments must be well determined and consistent. For good calibrations, the signal-to-noise ratios must be excellent and the angular sizes of the calibrators well determined. This may restrict calibrators to Solar System objects. Since Herschel cannot observe sources close to the Sun, and the PACS and SPIRE detectors will saturate when observing intense sources, the calibrations may have to be done using the emission from asteroids such as Vesta, Ceres, moons of outer planets, or smaller planets such as Uranus, Neptune or Pluto.

The datsets that will be produced by Herschel and ALMA will be so large that there may have to be special procedures and archives to ensure optimal synergy. The analysis and comparisons with models will have to be made on an automatic basis without a great deal of human intervention. Such computer analysis programmes have been developed by Schöier et al., 2005, for example, but these must be further developed to efficiently accommodate the very large data cubes that will be produced by ALMA and Herschel in the near future.

The following sections discuss the factors to be considered in detailed plans.

## 8.1 Allocation of Time

The Herschel team must take the lead in initiating projects, since the Herschel satellite has a relatively short lifetime. For any synergy, Herschel's role is twofold. The PACS and SPIRE bolometer cameras can be used for a rapid coverage of large regions of the sky, so these instruments could provide finding surveys for ALMA. On average for extragalactic sources, Herschel is limited to sources with redshifts $z < 3$ because of the combination of the shape of the thermal spectrum and the k-correction. Since ALMA measures dust emission longward of the peak of the Planck curve, such limits are much less severe. In our galaxy there are a large number of intense sources. Thus, the PACS and SPIRE cameras can survey large areas quickly. For both continuum and spectral lines, ALMA would image selected sources with higher angular resolution. For Herschel HIFI, the most interesting measurements are of the water vapour molecule. These measurements will be restricted to a limited number of sources since HIFI is a single pixel instrument. Once again, Herschel measurements will be first in time. Later ALMA would be used to provide images of water vapour lines using ALMA Band 5 receivers. With other receiver bands ALMA could be used to image dust continuum and spectral lines of other species.

Simultaneous observations are needed for rapidly time variable sources, such as comets. This will also be the case for continuum measurements of extragalactic variable sources such as active galactic nuclei or quasars. Simultaneous measurements seem to be less critical for time variable water vapour sources.

The full ALMA will be in operation toward the end of Herschel's life, so the data rights for Herschel may have expired. There is little need to have simultaneous measurement periods except for the cases as noted above. It is possible that Herschel will detect interesting but heavily obscured objects for which there can be no redshift measurements. Here there would have to be a need for quick responses from the ALMA Observatory to measure redshifts using CO rotational lines.

## 8.2  Supporting Observations

The best approach is to carry out lower angular resolution preparatory surveys of the GOODS, HDF-S and/or CDFS and galactic fields in the millimetre, sub-mm or far infrared. One possibility would be very deep integrations for extragalactic sources with the 12 m APEX antenna and the LABOCA bolometer camera. Larger single dishes with large bolometer cameras would provide better sensitivity, higher mapping speed, and also lower source confusion levels. Possible choices might be the James Clerk Maxwell (JCMT) with the SCUBA-2 camera. It is possible that the Large Millimeter Telescope (LMT), now under construction, will be in operation soon. The LMT with a bolometer camera would provide an excellent finding survey for ALMA. Further in the future, the use of the Extremely Large Telescope, ELT, in the sub-mm range would also be a possible choice if the ELT has horizon limits similar to ALMA's. For galactic sources, one could use the results of appropriate Spitzer Legacy programmes to image protostars or debris disks.

## 8.3  Supporting Data

There is a need to establish molecular collision rates, radiative transfer algorithms and chemical reaction rates. For some molecules this is included in the EC Framework Programme 6 *The Molecular Universe*. The scope of this programme is limited, and much more is needed. The *Molecular Universe Programme* should serve as a model for future undertakings.

One of the most important items for ALMA and Herschel is a completely automated form of data reduction. There will be very large three-dimensional data cubes of large collections of spectral line images (ie, two coordinates and velocity) or large two-dimensional broadband images from ALMA. With combined ALMA and Herschel datasets, the broadband results also form a data cube, with the third dimension being wavelength. These datasets may have sizes of more than a few million data-points; this precludes any detailed analysis using simple position-by-position analysis. The obvious step is to compare such datasets with models in an automated way. Clearly for specific regions one would start with general geometries, or pre-existing models based on lower angular resolution data, iterate the model until convergence, and then check whether the model is realistic in some sense. There is no general data reduction process available now, nor is there financial support for such a development. This will have to be the next step in the analysis of combined Herschel-ALMA datasets.

## 8.4 Calibration

To compare the Herschel and ALMA data, one must have accurate calibration. This will require a special measurement and source modelling programme. For ALMA, a science specification is to have absolute amplitude errors of 3% in the mm and 5% in the sub-mm. Thus any calibration source models must have accuracies of better than 1%.

The limitations of ALMA and Herschel are very different. The PACS and SPIRE detectors are easily saturated. Thus it is likely that one would use the fainter, smaller outer planets or asteroids. Since the bolometer bandwidths are much larger than the ALMA bandwidths, common calibrations require a very accurate model of the sources including possible time variations.

For Herschel HIFI and ALMA, there are common regions of the spectrum. However, the choice of spectral line calibration sources is more complex than with PACS or SPIRE. One may have to use either very small spectral line sources that are unresolved by ALMA, but bright enough to allow Herschel measurements, or (more likely) carry out a complete imaging of slightly extended calibration sources with ALMA. The first choice is limited by signal-to-noise ratios, while the second requires a larger investment in observing time with ALMA.

## 8.5 Data Archives

ALMA data alone and Herschel data alone will each be a great step forward. Combined ALMA-Herschel datasets will be a tremendous advance. Allowing open access to data of general interest by means of Virtual Observatory (VO) is the only way for an optimum exploitation of Herschel data and subsequent ALMA observations. ALMA and Herschel data must be VO compliant. There must be a clear policy of data rights since there will be only a short time interval in which common measurements are possible.

## 8.6 Action Items for the Organisations

*ESA:* There is a need to plan for Herschel surveys for regions within the ALMA elevation range. The calibration measurements must be carried out so that there are secure comparisons with ALMA continuum and spectral line data. The 75 µm band of PACS Herschel is needed to determine SEDs. These surveys will be unique; this wavelength is a fundamental input for the discrimination between AGN phenomena and starbursts in galaxies.

For an efficient synergy, ESA should devote a significant part of Herschel time to Legacy projects, ie, projects of wide interest for the community, starting soon after the science verification phase or very early in Herschel's lifetime. This would have to be the case for Herschel surveys of galactic and extragalactic sources, in continuum and spectroscopy. A targeted survey of Spitzer sources for ALMA follow-ups is needed. Such a survey would include debris disks, protostars, and galaxies. ESA must provide wide access to the data, calibration files and to a pipeline for quick looks as well as a package for complete data reduction. Ideally the pipeline products would be publication quality images.

***ESO:*** There is a need to carry out finding surveys for extragalactic/galactic sources using single dishes such as APEX with the LABOCA bolometer camera system.

For the time when both instruments are available, an efficient approach would involve a process where ESO reacts quickly to the need for follow-ups to Herschel data. For ALMA it would be useful to allocate observing time as soon as possible to measure variable sources, newly discovered sources, peculiar objects or in general to perform follow-ups of selected fields both in line and continuum that are as complete as possible.

***Both Organisations:*** The most pressing need is to coordinate large surveys that are planned in Herschel guaranteed time. There should be a series of dedicated conferences/workshops once the guaranteed time plans for Herschel are complete. These workshops should continue during the lifetime of Herschel. There may be difficulties in arranging for any coordination that involves large blocks of ALMA observing time, since the policies and organisation of ALMA Time Allocation Committees are in the discussion phase.

# Abbreviations

ACA — ALMA Compact Array
AGB — asymptotic giant branch
AGN — active galactic nucleus
ALMA — Atacama Large Millimeter Array
APEX — Atacama Pathfinder Experiment
ASTRO-F — now named AKARI, a Japanese infrared satellite
CDFS — Chandra Deep Field South
CIRB — cosmic infrared background
COB — cosmic optical background
DRSP — Design Reference Science Plan
ELT — Extremely large Telescope
ESA — European Space Agency
ESO — European Southern Observatory
FIR — far infrared
FP6 — Framework Program 6 (of the EC)
FWHP — Full Width to Half Power
GLIMPSE — the Galactic Legacy Infrared mid-plane Survey Extraordinaire
GMC — Giant Molecular Cloud
GOODS — Great Observatories Deep Survey
HDF-S — Hubble Deep Field South
HIFI — Heterodyne Instrument for the Far Infrared
HST — Hubble Space Telescope
IF — intermediate frequency
IRAM — Institut de Radio Astronomie Millimétrique
ISM — interstellar medium
ISOCAM — Infrared Space Observatory (near infrared) Camera
JCMT — James Clerk Maxwell Telescope
LABOCA — Large Bolometer Camera
LMT — Large Millimeter Telescope
LSB — lower sideband
MAMBO — Max Planck Millimeter Bolometer
MIPS — Multiband Imaging Photometer for SIRTF (now Spitzer)
MIR — mid infrared
ODIN — International Astronomy and Aeronomy Satellite
PACS — Photodetector Array Camera & Spectrometer
pwv — precipitable water vapor
rms — root mean square
SCUBA — Submillimeter Common User Bolometer Array
SED — Spectral Energy Distribution
SEST — Swedish Submillimeter Telescope
SIS — Superconductor Insulator Superconductor (a low noise microwave mixer)
SNR — supernova remnant

SWAS — Submillimeter Wave Astronomy Satellite
VO — Virtual Observatory